\documentclass[fleqn,10pt]{wlscirep}
\usepackage[utf8]{inputenc}
\usepackage[T1]{fontenc}
\usepackage[utf8]{inputenc}
\usepackage{CJKutf8}
\usepackage{amssymb}
\usepackage{amsthm}
\usepackage{amsmath,amssymb}
\usepackage{autobreak}
\usepackage{algpseudocode}
\usepackage{algorithm}
\usepackage{url}
\usepackage{graphicx}
\usepackage{float}
\usepackage{comment}
\usepackage{type1cm}
\usepackage{mathrsfs}
\usepackage{mathtools}
\usepackage{url}
\usepackage{booktabs}
\usepackage{braket}

\usepackage{bm}

\usepackage[switch]{lineno}

\usepackage{color}
\allowdisplaybreaks[1]

\newcommand\HL[1]{{\color{black}#1}}

\newcommand\HLL[1]{{\color{black}#1}}

\newcommand\HLLL[1]{{\color{black}#1}}

\newcommand\HLLLL[1]{{\color{black}#1}}

\newcommand\HLJPCOMPX[1]{{\color{black}#1}}

\title{Effects of topological structure and destination selection strategies on agent dynamics in complex networks}

\author[1,     a]{\textasteriskcentered Satori Tsuzuki}
\author[2,3,   b]{Daichi Yanagisawa}
\author[1,2,   c]{Eri Itoh}
\author[2,1,3, d]{Katsuhiro Nishinari}

\affil[1]{Research Center for Advanced Science and Technology, The University of Tokyo, 4-6-1, Komaba, Meguro-ku, Tokyo 153-8904, Japan}
\affil[2]{Department of Aeronautics and Astronautics, School of Engineering, The University of Tokyo, 7-3-1, Hongo, Bunkyo-ku, Tokyo, 113-8656, Japan}
\affil[3]{Mobility Innovation Collaborative Research Organization, The University of Tokyo, 5-1-5, Kashiwanoha, Kashiwa-shi, Chiba, 277-8574, Japan}

\affil[a]{tsuzukisatori@g.ecc.u-tokyo.ac.jp}
\affil[b]{tDaichi@mail.ecc.u-tokyo.ac.jp}
\affil[c]{eriitoh@g.ecc.u-tokyo.ac.jp}
\affil[d]{tknishi@mail.ecc.u-tokyo.ac.jp}

\begin{abstract}
We \HLLLL{analyzed} agent \HLLLL{behavior} in complex networks: Barab\'{a}si-Albert (BA), Erdos-R\'{e}nyi (ER), and Watts-Strogatz (WS) models under the following rules: agents (a) randomly select a destination among adjacent nodes; (b) exclude the most congested adjacent node as a potential destination and randomly select a destination among the remaining nodes; or (c) select the sparsest adjacent node as a destination. 
\HLLLL{We focused on small complex networks with node degrees ranging from zero to a maximum of approximately 20 to study agent behavior in traffic and transportation networks.} We measured the hunting rate, that is, the rate of change of agent amounts in each node per unit of time, and the imbalance of agent distribution among nodes. Our \HLLLL{simulation} study reveals that the topological structure of a network precisely determines agent distribution when agents perform full random walks; however, their destination selections alter the agent distribution. Notably, rule (c) makes hunting and imbalance rates significantly high compared with random walk cases (a) and (b), irrespective of network types, when the network has a high degree and \HLLL{high activity rate}. Compared with the full random walk in (a), (b) increases the hunting rate while decreasing the imbalance rate when activity is low; however, both increase when activity is high. These characteristics exhibit slight periodic undulations over time. 
\HLJPCOMPX{Furthermore, our analysis shows that in the BA, ER, and WS network models, the hunting rate decreases and the imbalance rate increases when the system disconnects randomly selected nodes in simulations where agents follow rules (a)--(c) and the network has the ability to disconnect nodes within a certain time of all time steps.}
\HLLL{Our findings can be applied to various applications related to agent dynamics in complex networks.}
\end{abstract}

\begin{document}

\flushbottom
\maketitle
%
%
\thispagestyle{empty}









\section{Introduction}
\HLLL{A complex network comprises numerous nodes connected by edges.}
Elucidating the effect of agent decision-making on complex networks is of utmost importance. 
In the real world, \HLLL{complex} networks exist on which agents travel, as exemplified by traffic transport systems~\cite{doi:10.1080/01441647.2013.848955}, socio-physical networks~\cite{Xu2011NonliDyn} such as crowds, virtual networks on the internet, and biological brain networks. Here, agents represent objects such as parcels, self-driven objects such as humans or vehicles, or the units of data or signals. 
In the case of agents' self-intelligence and that of receiving instructions from a manager, agent destination selection strategies can significantly affect the overall network behavior. Particularly, avoiding congestion~\cite{JABBARPOUR2014303, ARNOTT1991309, 6666580} can result in different agent mobilities. An example is provided in our previous study~\cite{Tsuzuki2022SciRep}, which explored the effect of agents avoidance congestion on agent mobility and distribution among multiple nodes of a star topology. We demonstrated the existence of an optimized number of nodes that resulted in the most uniform agent distribution. Regarding the travel efficiency of the star topology, we found that agents conducting congestion avoidance linearizes the travel time for all agents visiting all nodes; however, the travel time increases exponentially with the number of nodes when they do not avoid congestion~\cite{Tsuzuki2022SciRep}. A star topology is one of the simplest graphs because it consists of a primary node and multiple secondary nodes. Complex networks are composed of numerous star topologies; the congestion avoidance behavior of agents may significantly impact agent dynamics on complex networks. 

\HLLL{In contrast}, the static structure of a network \HLLL{often} affects the agent dynamics in the network, and such a mechanism may be observed in the human brain. \HLLL{Several} studies~\cite{doi:10.1126/science.1238411, Avena-Koenigsberger2018} have reported that the topological structure of brain networks forms interaction and signaling patterns among neurons and brain regions. The resulting communication mechanism is critical to the brain; however, their detailed mechanism remains unclear. \HLJPCOMPX{Studies on synchronization and consensus problems in multi-agent communication have also been reported~\cite{ShangYi-Lun_2010, 10.1063/1.4976959}}.
Other scientists have reported the effects of network structures on information spreading~\cite{liu2015events, NIE2021127098}. \HLLL{In addition}, other related studies have utilized random walk models~\cite{10.1371/journal.pcbi.1006833, 10.1063/1.4939837, weng2023multiple, PhysRevE.98.052302, MASUDA20171, PhysRevE.75.016102} to simulate the dynamics of crowds and neurotransmitters, such as dopamine, or to search for the \HLLL{topologies} of \HLLL{complex} networks.
Clarifying the impact of destination selection strategies on agent dynamics in complex networks and the effect of static network structure on agent dynamics in complex networks is \HLLL{significant} for \HLLL{various} applications.

\HLLL{This study investigates} the agent dynamics on three complex networks: Barab\'{a}si-Albert (BA)~\cite{doi:10.1126/science.286.5439.509, albert2002statistical, doi:10.1126/science.1173299}, Erdos-R\'{e}nyi (ER)~\cite{erdHos1960evolution}, and Watts-Strogatz (WS)~\cite{Watts1998, Barrat2000} models under the following rules: agents (a) randomly select a destination among adjacent nodes, (b) exclude the most congested node among neighboring nodes as a destination and randomly select a destination among the remaining nodes, or (c) select the sparsest adjacent node as a destination.
The BA, ER, and WS models are mathematical models designed in the network science field to reproduce the behavior and properties of real-world complex networks. 
\HLLL{First, the BA model is a network model in which nodes are preferentially added to the existing nodes with larger connections, leading to the formation of hubs and a power-law degree distribution. Examples of the BA models vary from the World Wide Web, air transport networks, and biological networks~\cite{Almaas2006}.}
\HLLL{Second,} the ER model is a random model that generates networks with a Poisson degree distribution, in which nodes are randomly connected with a fixed probability. The resulting network has fewer clusters and a homogeneous degree distribution. 
\HLLL{Finally,} the WS model has a high clustering coefficient and short path lengths, \HLLL{reproducing the small-world nature of social networks}~\cite{10.2307/2786545}. A set of two nodes adjacent to a node \HLLL{is} likely to be connected to each other (``small-world''). 

As evaluation criteria for agent dynamics in complex networks, we introduce the following two characteristics: the hunting rate, defined as the rate of change of the agent amounts in each node per unit time step, and the imbalance of the agent distribution among nodes. As explained previously, we measured these characteristics using Rules (a)--(c). These rules primarily assume decision-making situations that we frequently \HLLL{observe in the real world} when \HLLL{agents move} from current locations to destinations. Compared with the full random walk scenario in Rule (a), avoiding the most congested site in Rule (b) is a typical human behavior observed in congested crowds. \HLLL{Congestion-avoidance behavior itself can also be observed in social insects such as ants and bees}~\cite{BLUM2005353, Frisch+1993}. Selecting the sparsest adjacent site as a destination would be \HLLL{an} evacuation behavior during an emergency. Hence, the primary application of this study is in \HLLL{collective} dynamics. However, Rules (a)--(c) \HLLL{may} also be applied to other control systems. In air-ground traffic systems, the air controller often instructs aircraft to avoid congested spots in taxiing~\cite{mazur2018simulation, Tsuzuki_2020, 9878328}. \HLJPCOMPX{In this way, the target scenarios under Rules (a)--(c) have great applicability to a wide range of network problems. However, our primary interest is in transportation networks and pedestrian networks, such as visitors to event venues, where the number of route branches would be at most about 20. Therefore, we focus on small complex networks with node degrees ranging from 0 to 20 to study agent behavior in transportation and pedestrian networks}
\HLLLL{As a method for this study}, we \HLL{developed a network of} cellular automata (CA)~\cite{10.5555/1102024, RevModPhys.55.601}, where each cell represents a node. Specifically, each agent on a node repeatedly moves to its adjacent nodes according to Rules (a)--(c), which is a straightforward but promising approach. Interestingly, in multivariate statistics, congestion-avoidance behavior is described as a mutual correlation between agents via congestion information~\cite{Tsuzuki2022SciRep}, where agents interact with each other. 
\HLJPCOMPX{Specifically, agents are prompted by the provided congestion information to congregate at less congested nodes while avoiding congested nodes. From an agent-agent interaction perspective, we see this situation as agents being attracted to other agents at less congested nodes while being repelled by agents at congested nodes; this is a kind of interaction among agents via the provided congestion information.} Hence, the target system is a multi-agent system. 
\HLJPCOMPX{However, computing the multiagent reinforcement learning (MARL) systems gets computationally extensive as the number of agents increases~\cite{9238072, 10.5555/3463952.3464013}. In addition, it is still under discussion how to reproduce the functionality of congestion avoidance in a probability transition function.}
\HLLLL{Therefore}, we performed CA simulations to investigate complex networks. Section~\ref{sec:targetsys} details the proposed methodology.

The remainder of this study is organized as follows. Section 2 describes the design variables of the target problem. In addition, we represent two characteristic values, hunting and imbalance rates, that is, the criteria for agent mobility in complex networks. Next, we describe the numerical conditions of the benchmark tests. Section 3 reports the results of the agent simulations in the aforementioned network models: the BA, ER, and WS models for Rules (a)--(c). Section 4 highlights the four significant findings of our results. (1) The degree distribution, that is, the topological structure of a network, precisely determines the agent distribution among nodes in the steady state when agents perform a fully random walk. \HLL{We demonstrated that the theory of a grand canonical ensemble in statistical physics describes this fact well.}
In addition, the agents' decisions to select their destination can alter the overall agent distribution. (2) Agents deciding to always select the sparsest adjacent node makes both the hunting and imbalance rates on a complex network significantly high compared with random-walk scenarios, regardless of the network type, particularly when the network has a high degree and activity rate. However, this is not necessarily true when the network has a significantly low degree. (3) The agents' decision to exclude the most congested node among neighboring nodes as a destination increases the hunting rate while decreasing the imbalance rate when the agent activity is low. Conversely, when agent activity is high, their decision increases both the hunting and imbalance rates, which is true for almost all cases of the ER and WS models owing to the relatively uniform degree distribution but not necessarily true for the BA model because of the localized distribution. (4) The hunting and imbalance rates show periodic undulations with respect to time. We discuss the application of these findings in real-world physical and biological networks. Finally, Section 5 summarizes this study.

\section{Methods}\label{sec:targetsys}
A network is represented as a graph consisting of nodes connected by edges. In this study, we focused on undirected graphs. We introduce three measures for complex networks. First, the degree represents the number of edges per node. The average degree ($k$) refers to the average degree value for all nodes. The path length indicates the number of minimum edges traced from one node to another. The average path length ($l$) represents the average path length of all nodes. In addition, we refer to a group of three nodes that are mutually connected to form a triangle, that is, the closed path with a length of two, as a cluster. We defined the clustering coefficient ($C$) as the ratio of the number of clusters to the total number of paths with a length of two. $k$, $l$, and $C$ are the key characteristics representing the spatial structure of complex networks. 

In this study, the target system is described as a set $(N_n, N_a, N_t, A_r, G(X))$, where $N_n$ represents the number of nodes, $N_a$ indicates the number of agents, and $N_t$ represents the number of timesteps in each simulation. $A_r$ indicates the probability of each agent hopping from one node to another connected node, and $A_{r}$ represents the activity rate of the agents. $G$ is a type of graph: a BA, ER, or WS model. The $X$ of G symbolically suggests a set of design variables for each graph. The BA and ER models have a single design variable in addition to variables $(N_n, N_a, N_t)$. 
Specifically, the BA model has a design variable $N_J$ that indicates the number of edges attached from a newly added node to the existing nodes in the creation process. The preference of an existing node to be connected to a newly-added node is given by their degrees; that is, the probability for the $i$th node to be selected as one side of $N_J$ new edges is provided by the ratio of the $i$th node's degree $k_i$ to the sum of the degrees of all the nodes: $k_{i}/\sum k_{i}$, which suggests that the node with higher degrees can be preferentially selected. This is the reason why the BA model has a so-called ``hub,'' a gathering of densely connected nodes. The resulting degree distribution is known to have the scale of $k^{-3}$, a power-law distribution; therefore, the BA network is called ``scale-free.'' The average path length $l$ and clustering coefficient $C$ have scales of ${\rm ln}(N_{n})/{\rm ln ln} (N_{n})$ and $N_{n}^{-0.75}$, respectively, in the BA model~\cite{10.1371/journal.pone.0013701}. 
In contrast, the ER model generates an edge between each pair of nodes with probability $P_r$. That is, each of the $N_{n}(N_{n}-1)/2$ combinations of $N_{n}$ nodes creates an edge with probability $P_{r}$. When $P_{r}=1$, each node has $N_{n}-1$ connections; conversely, if $P_{r}$ is sufficiently small, isolated nodes with no connection to other nodes emerge. Because each pair of nodes has an edge with probability $P_r$ and does not have it with $1-P_{r}$, the probability of the existing $M$ edges among $N_{n}$ nodes obeys the binominal distribution. Additionally, the average path length $l$ and clustering coefficient $C$ of the ER model have scales of ${\rm ln}(N_{n})/{\rm ln} (k)$ and $k/N_{n}$, respectively, ~\cite{10.1371/journal.pone.0013701}. The generation of complex networks with isolated nodes should be avoided because this study focuses on the agent dynamics in complex networks. We examined the connectivity of the generated static networks for different values of $P_{r}$ between zero and one before performing agent simulations. We determined the suitable parameter range of $P_{r}$ to be between $0.4$ and $1.0$ for the simulations described in Section~\ref{sec:result}. 

Unlike the BA and ER models, the WS model has two design variables: the number of nodes connected to each node in the initial state $N_J$ and the probability of rewiring edges $P_r$. During the creation process, all nodes are arranged in a ring at regular intervals in the initial state. Each node has an edge to its neighboring nodes. Here, the range of neighbors depends on the size of $N_J$; when $N_J = 2$, each node has edges only to the closest nodes in addition to the adjacent nodes on the ring. In contrast, it has four edges to the closest and second-closest nodes when $N_J = 4$. As the value of $N_J$ increases, neighboring nodes become mutually connected. Thus, the clustering coefficient $C$ increases with $N_J$. By rewiring the edges among the nodes with probability $P_r$, the clusters collapse; as $P_r$ increases, $C$ decreases. The average path length $ l $ decreases rapidly owing to the rewiring process, which connects a node with the non-neighboring nodes. This rapidly decreases the average path length, and eventually, the network approaches a random network structure. When a network has a small $l$ while maintaining a high clustering coefficient $C$, the network is called a ``small-world'' network because the neighboring nodes of a node are likely to be connected. That is, a person's friends are likely to be friends. As for the remaining key characteristics, the average path length $l$ and clustering coefficient $C$ of the WS model have scales of $\frac{N_{n}}{k}\cdot f(P_{r} k N_{n})$  and $\frac{3(k-2)}{4(k-1)}(1-P_r)^{3}$, respectively~\cite{10.1371/journal.pone.0013701}, where $f(x)$ represents a function that becomes constant when $x \ll 1$ and becomes ${\rm ln}(x)/x$ when $x \gg 1$. In summary, the independent variables of the BA model are $(N_n, N_a, N_t, A_r, N_J)$, those of the ER model are $(N_n, N_a, N_t, A_r, P_r)$, and those of the WS model are $(N_n, N_a, N_t, A_r, N_J, P_r)$. The physical significances of $N_J$ and $P_r$ differ according to the model. As described previously, for the ER model, $P_r$ is confined between $0.4$ and $1.0$; otherwise, $0$ and $1$. We investigated target systems with different values of $N_J$ between $1$ and $10$, assuming the number of junctions or branches of real-world traffic networks on which the agents travel. Activity rate $A_{r}$ was set from $0.1$ to $0.9$ at the initial state. The method for determining the remaining parameters $(N_{n}, N_{a}, N_{t})$ is described below.

\begin{figure*}[t]
\vspace{-22.5cm}
\includegraphics[width=4.6\textwidth, clip, bb= 4 4 3176 1111]{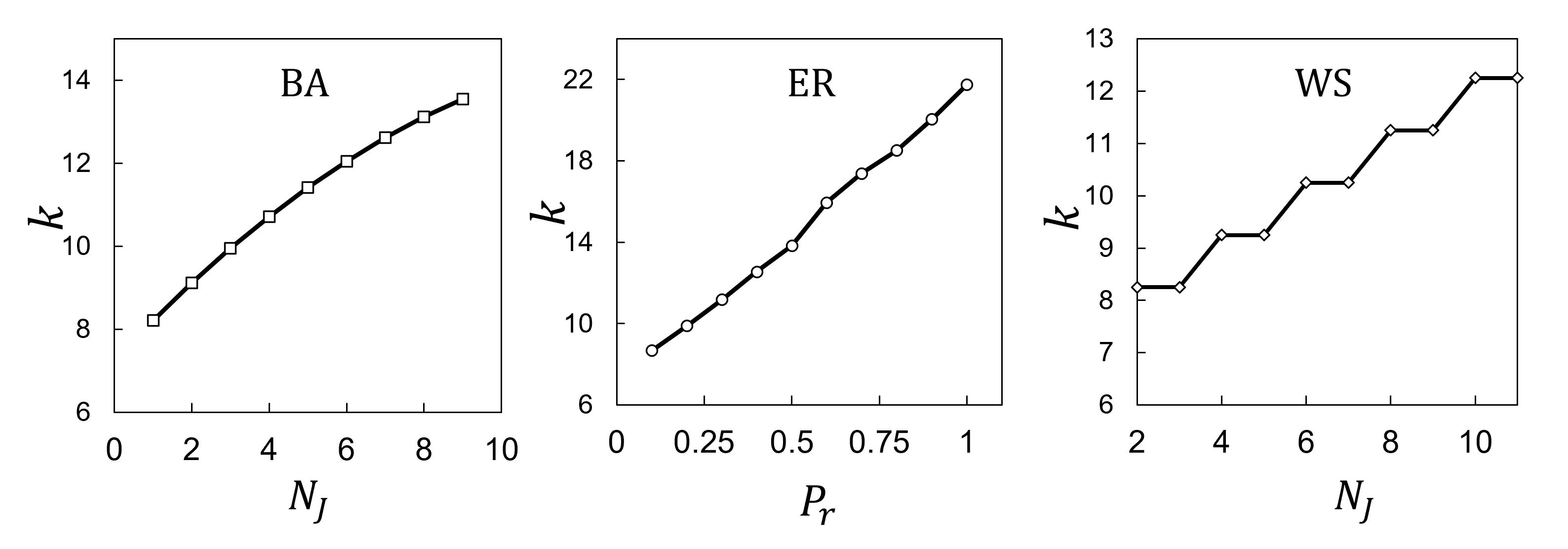}
\caption{Dependence of average degree $k$ on the respective design variables $N_{J}$, $P_{r}$, and $N_J$ for the BA, ER, and WS models, respectively.}
\label{fig:Figure1}
\end{figure*}

We used NetworkX~\cite{SciPyProceedings_11}, a most well-established Python package for complex networks. Specifically, we implemented the BA network using {\sf \small barabasi\_albert\_graph}, the ER network using {\sf \small erdos\_renyi\_graph}, and the WS network using {\sf \small watts\_strogatz}\\{\sf \small \_graph}. Parameter $N_J$ of the BA model corresponds to the input parameter ``$m$'' of {\sf \small barabasi\_albert\_graph}. Because they determine the number of edges attached from a newly added node to the existing nodes, the BA model has a high degree as $N_J$ increases. Similarly, parameter $P_r$ of the ER model corresponds to the input parameter ``$p$'' of {\sf \small erdos\_renyi\_graph} that defines the probability of creating an edge between each pair of nodes; as $P_r$ increases, the ER model has a high degree. Additionally, parameters $N_J$ and $P_r$ correspond to the ``$k$'' and ``$p$'' of {\sf \small watts\_strogatz\_graph}, where $k$ represents the number of nodes connected to each node at the initial state, and ``$p$'' indicates the probability of rewiring edges. Thus, as $N_J$ increased, the WS model exhibits a higher degree. Figure~\ref{fig:Figure1} shows the dependence of the average degree $k$ on the design variables $N_{J}$ for BA, $P_{r}$ for ER, and $N_J$ for WS. We can confirm that $N_J$ increases for the BA or WS models and $P_r$ increases for the ER model, and that each model has a higher degree. Reasonably, the WS model shows a step function because it connects $N_{J}-1$ neighbors if $N_{J}$ is odd~\cite{Watts1998, SciPyProceedings_11}. In addition, we confirm that the average degree $k$ is independent of $P_{r}$ in the WS model.
\HLJPCOMPX{To be clear, the fact that the average degree $k$ increases as the parameter $N_J$ or $P_r$ increases does not mean that the clustering coefficient $C$ also increases as $N_J$ or $P_r$ increases. In fact, related studies suggest that the magnitude of the clustering coefficients asymptotically converges to the average clustering coefficient in certain cases~\cite{LI2017350}. In any case, we have confirmed the average degree $k$ increases as $N_J$ or $P_r$ increases within the focused parameter range.}

\HLL{As mentioned in the Introduction, we investigate the agent dynamics on three complex networks under the following rules: agents (a) randomly select a destination among adjacent nodes; (b) exclude the most congested node among neighboring nodes as a destination and randomly select a destination among the remaining nodes, or (c) select the sparsest adjacent node as a destination. The mathematical expressions for Rules (a)--(c) on the $i$th node are as follows:}
\HLL{
\begin{eqnarray}
D^{(i)}_{s}(X),~~X~\coloneqq~
\begin{cases}
~\{ x \in N_{d}^{(i)}~|~x = ~{\rm rand}~(n_{a}^{1},n_{a}^{2},\cdots,n_{a}^{k_{i}})~\} & {\rm if~~Rule~(a)},\\
~\{ x \in N_{d}^{(i)}~|~x = ~\overline{{\rm max}}~(n_{a}^{1},n_{a}^{2},\cdots,n_{a}^{k_{i}})~\}& {\rm if~~Rule~(b)},\\
~\{ x \in N_{d}^{(i)}~|~x = ~{\rm min}~(n_{a}^{1},n_{a}^{2},\cdots,n_{a}^{k_{i}})~\} & {\rm if~~Rule~(c)},\\
\end{cases}
\end{eqnarray}
where $N^{(i)}_{d}$ represents the set of all nodes indexed from the viewpoint of the $i$th node, and $k_{i}$ represents the $i$th node degree, that is, the number of nodes connected to the $i$th node. $n^{j}_{a}$ indicates the number of agents in the $j$th neighboring node. The function $\rm \sf rand(\cdots)$ returns one of $\{n^{1}_a, n^{2}_a, \cdots, n^{k_i}_{a}\}$ at random; $X$ has only one selected element in this case. The function $\rm \sf \overline{max}(\cdots)$ returns the complement of the function $\rm \sf max(\cdots)$ that returns the maximum value among $\{n^{1}_a, n^{2}_a, \cdots, n^{k_i}_{a}\}$; in this case, $X$ usually has multiple elements. The function $\rm \sf min (\cdots)$ returns the minimum value among $\{n^{1}_a, n^{2}_a, \cdots, n^{k_i}_{a}\}$; $X$ has multiple elements when two or more of $\{n^{1}_a, n^{2}_a, \cdots, n^{k_i}_{a}\}$ have the same minimum value. $D^{(i)}_{s}(\cdots)$ is a function that converts $n^{l}_a$ to parameter $l$, a local index of the neighboring node that will be selected as a destination, after randomly selecting one element in $X$ when $X$ has multiple elements.}

We introduce the following two physical quantities as indicators of agent dynamics in a complex network: the hunting rate $H$, which is the change rate of agent counts in each node per unit time step, and imbalance rate $L$, which is the non-uniformity of the agent distribution among nodes. Using parameters $N_{n}$ and $N_{a}$, the hunting rate $H$ and imbalance rate $L$ are expressed in the following forms:
\begin{eqnarray}
H &\coloneqq& \frac{1}{N_{a}} \sum_{i=1}^{N_{n}} \biggl | {N_{n}}_{(m)}^{(i)} - {N_{n}}_{(m-1)}^{(i)} \biggr |, \label{eq:defparaH} \\
L &\coloneqq& \frac{1}{N_{n}} \sum_{i=1}^{N_{n}} \biggl | {N_{n}}_{(m)}^{(i)} - {N_{n}}^{avr} \biggr |, \label{eq:defparaL} 
\end{eqnarray}
where $N_{n}^{avr} = N_{n}/N_{a}$ is the average number of agents per node. ${N_{n}}_{(m)}^{(i)}$ represents the number of agents on the $i$th node at the $m$th time step. Equation (\ref{eq:defparaH}) shows that parameter $H$ increases as the change rate of the number of agents per unit of time increases, suggesting that $H$ is an indicator of the risk that agents thrust into a node in a short time. We can say that agent mobility in a complex network is moderate when the value of $H$ is low in the stationary state. Conversely, when the system exhibits high values of $H$, the agents become rapidly mobile. We introduced $1/N_{a}$ into Eq.~(\ref{eq:defparaH}) for normalization based on the problem size. For Eq.~(\ref{eq:defparaL}), parameter $L$ increases as the deviation of the number of agents from the average number of agents per node increases, which indicates that the agent distribution among nodes becomes uniform as $L$ decreases, and the agent distribution becomes non-uniform as $L$ increases. 
\HLL{In addition, we adopted the parallel update scheme for the time integration in which the states of all nodes are simultaneously updated at every time step.}

\HLJPCOMPX{Originally, ``hunting'' was used as an electrical engineering term referring to the phenomenon of unexpected oscillations in the time direction of a characteristic value in an electrical system, such as a power system~\cite{WU2000597}, of which a motor voltage disturbance is a typical example. The term is now being used in other fields and is becoming popular to refer to a phenomenon in which a certain characteristic of a complex system causes unexpected oscillations in the time direction. We refer to this as the intensity of variation of the agents at each node per unit of time in complex networks. In a real pedestrian network, the greater the number of agents moving along a given path in the network in the amount per unit of time, the more they can collide with each other, leading to an increase in accidents such as crowd avalanches in the most congested case. In this sense, parameter $H$ in Eq.~(\ref{eq:defparaH}) can be said to be an objective metric that corresponds to the risk measure for crowd accidents in such real networks. On the other hand, the imbalance rate $L$ in Eq.~(\ref{eq:defparaL}) is a measure of how much an agent distribution per unit time deviates from the case where agents are evenly distributed at each node. We are interested in how much $L$ changes when the rule is changed rather than the value itself. For example, changing Rule (a) to Rule (b) causes agents to avoid the most congested nodes. Each agent is a free agent under Rule (a), but under Rule (b), it becomes a decision entity that wants to avoid the most congested node. If we approximately consider the behavior of agents visiting an event site consisting of multiple booths as random walks, the variation of $L$ when we change Rule (a) to Rule (b) may serve as an indicator of how people's collective behavior changes when they are provided with congestion information and change from randomly wandering free agents to decision-making entities. Meanwhile, Rule (c), where agents always choose the most sparse node, reflects the tendency of pedestrians to avoid congestion during evacuations. We also know from our experience that we often avoid congested places when visiting amusement parks, shopping malls, or attending multiple events. Compared to Rule (b), the will to avoid congestion in Rule (c) is stronger because the agent in Rule (b) only avoids the most congested node and has the right to choose one of the other neighbors, whereas the agent in Rule (c) always makes a coercive and exclusive decision to choose the most vacant node. By comparing Rule (b) and Rule (c), we can see how the different degree of congestion avoidance decision manifests itself in the collective dynamics of the agents. In this way, Rules (a)--(c) are all associated with real-world scenarios. 

Many network properties discussed in previous studies are mainly related to spatial structures such as clustering coefficients and fractal dimension~\cite{Balaban2018, Yang2021, Xiao2021}. For multi-agent dynamics where agents perform mutual interactions such as attraction or repulsion, there is a study that presents statistical physical metrics, including self-organization and complexity~\cite{Koorehdavoudi2016}. In contrast, our metrics in Eqs.~(\ref{eq:defparaH}) and (\ref{eq:defparaL}) are oriented toward estimating agent mobility and distribution in a social system where people with self-intelligence of select destinations, such as visitors to an event site.}

First, we measured the values of $H$ and $L$ averaged over the elapsed time step for several cases of $(N_n, N_a)$ as a preliminary test. 
\HLJPCOMPX{This is to identify the effect of interactions between agents due to congestion avoidance behavior on the long-term evolution of these two metrics.}
Consequently, we found that $H$ and $L$ reached a steady state shortly after the simulation started in all cases of $(N_n, N_a)$, except that the time step required for the system to reach a steady state minorly differs in each case. 
Figure~\ref{fig:Figure4}(A) shows the dependencies of $H$ on the time steps for $(N_n, N_a)= (30, 2000)$ for the BA model with different values of $N_J$, as shown in Fig.~\ref{fig:Figure5}(A) for the ER model for different values of $P_r$, and Figs.~\ref{fig:Figure6}(A), ~\ref{fig:Figure7}(A), and ~\ref{fig:Figure8}(A) for the WS models with different values of $N_J$ and probabilities $P_{r}=0.1$, $P_{r}=0.3$, and $P_{r}=0.9$, respectively. Each column (A) of Figs.~\ref{fig:Figure4}--\ref{fig:Figure8} displays the cases of Rules (a), (b), and (c) from the left. \HLL{The same observations were made for $L$, as displayed in each column (C) of the corresponding figures for each model}. These results confirm that the value of $N_t$ required to reach a steady state can be specified as $(N_n, N_a)$, which is approximately 600 for $(N_n, N_a) = (30, 2000)$. 
\HLJPCOMPX{We also confirmed that the coefficients of variation for the time series data for the variation of the averaged values of $H$ and $L$ in a steady state area shown in (A) and (C) of Figs.~\ref{fig:Figure4}-\ref{fig:Figure8} were less than 0.5 percent.}
Accordingly, we start our analysis with the set $(N_{n}, N_{a}, N_{t}) = (30, 2000, 600)$ and then discuss the extendibility of the obtained results to other cases of $(N_{n}, N_{a})$.

\begin{figure*}[t]
\vspace{-48.5cm}
\includegraphics[width=4.5\textwidth, clip, bb= 4 4 4530 3490]{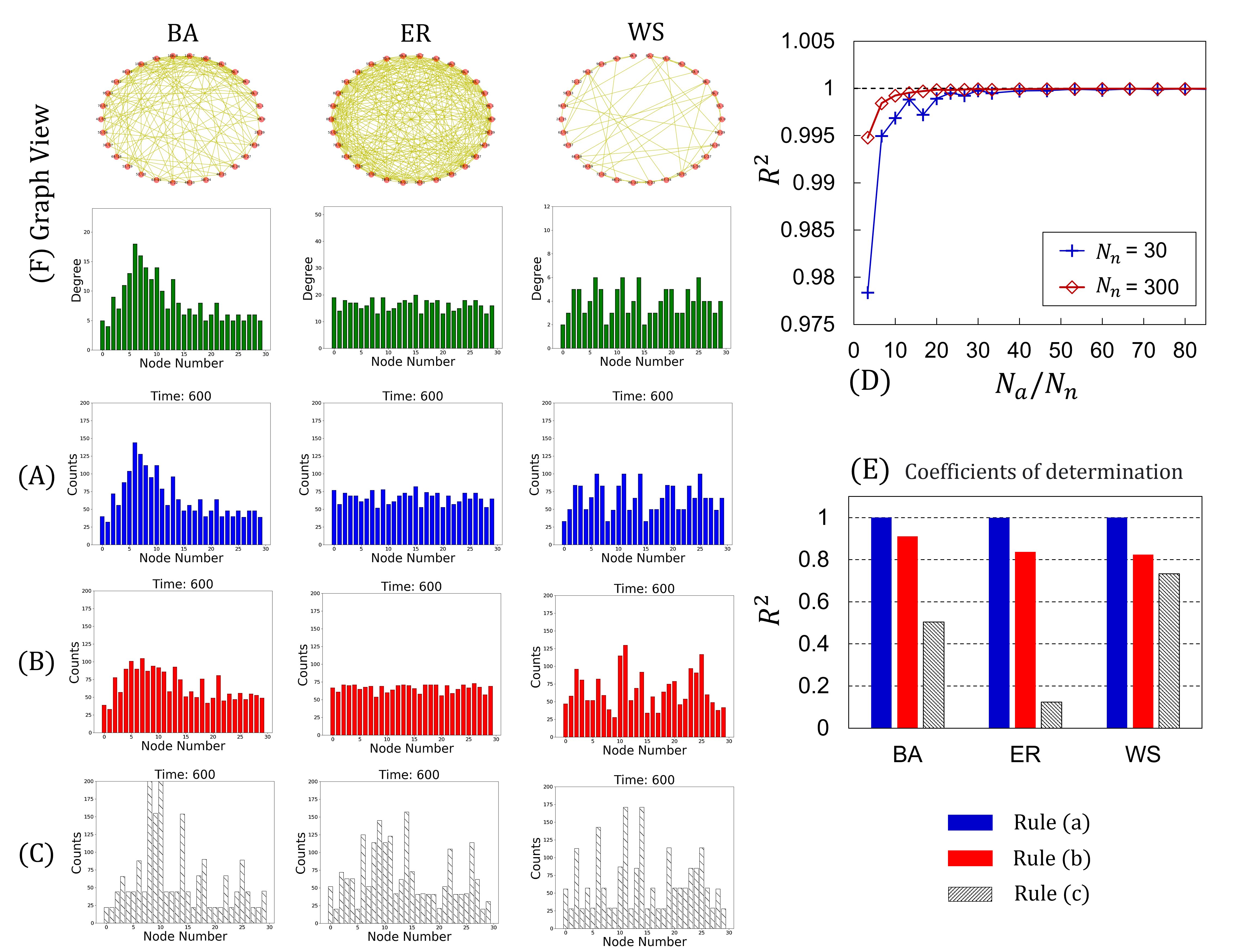}
\caption{(A) Agent distribution at a steady state for Rule (a); (B) agent distribution at a steady state for Rule (b); (C) agent distribution at a steady state for Rule (c); (D) relationship between the coefficient of determination and $N_{a}/N_{n}$ ratio, in both small and large cases of $N_{n}$; (E) comparison of the coefficient of determination for Rules (a)--(c) in the BA, ER, and WS models; (F) graphs generated in the three network models (top), along with their corresponding degree distributions (bottom).}
\label{fig:Figure2}
\end{figure*}

\begin{figure*}[t]
\vspace{-34.5cm}
\hspace{+1.5cm}
\includegraphics[width=3.5\textwidth, clip, bb= 4 4 1792 1258]{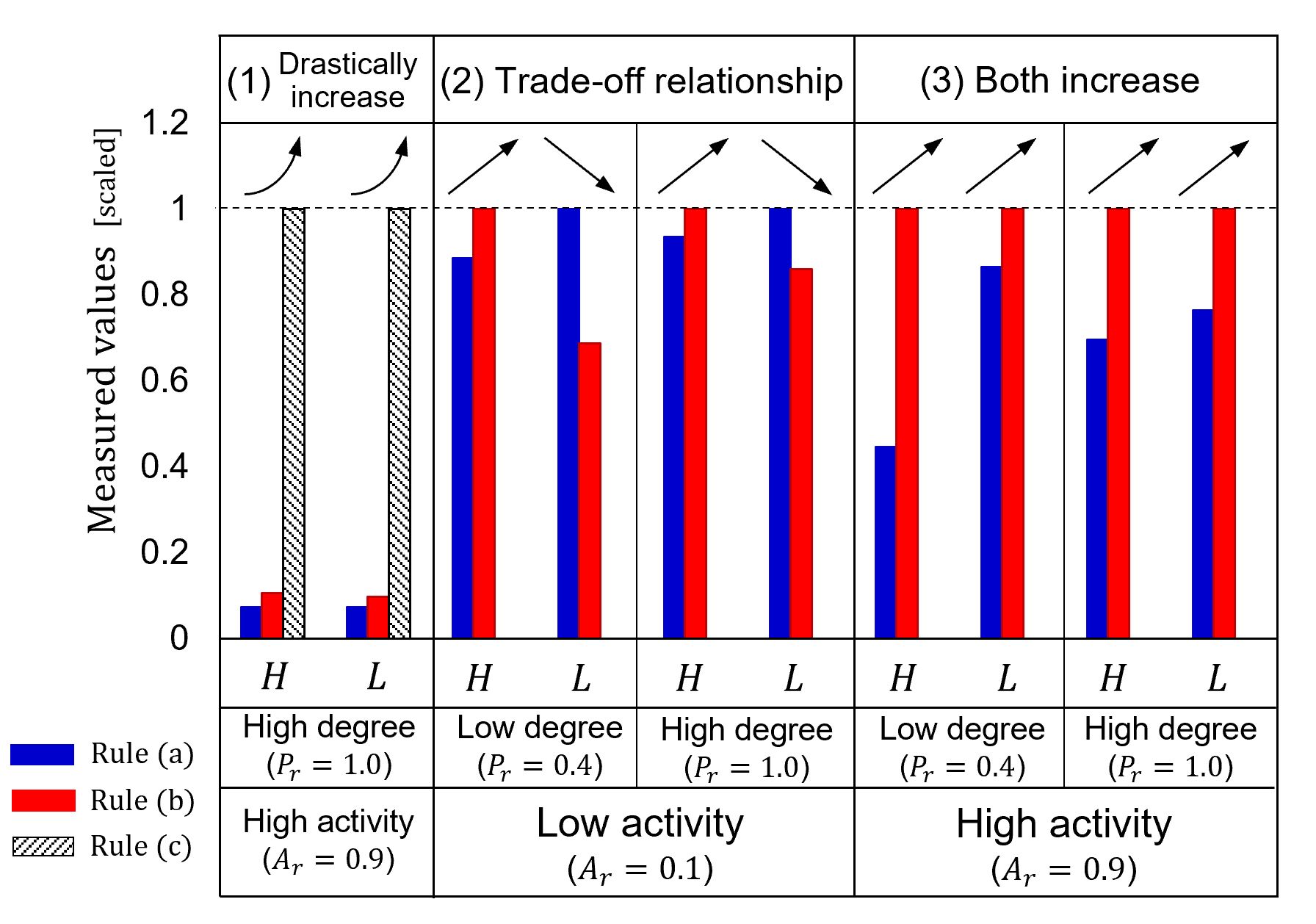}
\caption{Summary of the observed relationships for the ER model.}
\label{fig:Figure3}
\end{figure*}

\section{Analysis}\label{sec:result} 
\HL{We equally distributed the agents among nodes at the initial state}. We then conducted agent simulations wherein agents walked around among nodes according to Rules (a)--(c) under the condition of $(N_n, N_a, N_t) = (30, 2000, 600)$ for the BA, ER, and WS models for different values of $N_J$ and $P_{r}$. Consequently, the agent and degree distributions were observed to agree exactly irrespective of the network type and design variables $N_J$ and $P_r$. Figure~\ref{fig:Figure2} shows example results of the agent simulations when setting $N_J$ to 0.5, $P_r$ to 0.5, and $(N_J, P_r) = (0.5,0.5)$ for the WS models. In Fig.~\ref{fig:Figure2}(F), the top row illustrates the generated network graphs, and the bottom row presents the corresponding degree distributions. In addition, Fig.~\ref{fig:Figure2}(A) shows the agent distribution in the steady state for Rule (a). Compared with (A) and (F), the agent and degree distributions were confirmed to match in all three network model cases. The blue bars in Fig.~\ref{fig:Figure2}(E) show the values of $R^{2}$, which is the coefficient of determination between the agent and degree distributions for Rule (a); the $R^{2}$ value becomes one with significant accuracy in all three models. 

Let us \HLL{elaborate} on the agreement between the agent and degree distributions using statistical mechanics. 
We assume that the agent distribution reaches an equilibrium state, implying that the flows of agents per node are equal. Next, we consider the $i$th node, which has a degree of $k_{i}$, that is, $k_{i}$ connected nodes. Upon reaching the equilibrium state, we can consider the $i$th node a thermodynamic system connected to multiple reservoirs with agents as particles flowing into/out of the system; the system is a grand canonical ensemble~\cite{Callen:450289}. Thus, we consider the system an absorption problem. Specifically, we regard the $i$th node with $k_{i}$ neighboring nodes as a node with $k_i$ adsorption sites. When a single agent (particle) flows into the $i$th node, one of the $k_i$ adsorption sites is filled with the particle. Assuming that the energy state of a node is proportional to the number of agents when the system has only agents, the energy state of the system can be represented as $E_{0}$. This assumption is valid because the state of a node is designated only by the number of agents in the target system. However, the total number of ways in which $k_{m}$ sites among the $k_{i}$ sites are filled with agents corresponds to the number of combinations of selecting $k_{m}$ sites from $k_{i}$ nodes, expressed as $C_{i} = {k_{i}!}/{ {k_{m}!} {(k_{i}-k_{m})!} }$. Because each state has the same energy $E_{0}$ multiplied by $k_{i}$ as $E_{0} k_{i}$, the resulting grand partition function is expressed as $\Xi = [1 + {\rm e}^{\beta(E_{0}+\mu)}]^{k_{i}}$, where $\beta$ and $\mu$ denote the inverse temperature and chemical potential of the node. Using a series of formulas for the grand partition function, the expected number of sites filled with agents, that is, the expected number of agents in the $i$th node, is obtained by the partial derivative of the natural logarithm of $\Xi$ scaled by $\beta^{-1}$ with respect to the chemical potential $\mu$ as follows:
\begin{eqnarray}
\braket{N_i}~:=~\frac{1}{\beta} \frac{\partial}{\partial \mu} {\rm ln} \Xi~=~k_i \frac{{\rm e}^{\beta(E_0+\mu)}}{1 + {\rm e}^{\beta(E_0+\mu)}}. 
\label{eq:grandcanonanop}
\end{eqnarray}
Here, all nodes have the same parameters $\beta$ and $\mu$ because the activity rates $A_{r}$ of all nodes are the same. Hence, the expected distribution of the number of agents is expressed as follows:
\begin{eqnarray}
\{N_{1}, N_{2}, \cdots, N_{N_n}\}~=~D \{k_{1}, k_{2}, \cdots, k_{N_n}\},~~~~D~\coloneqq~\frac{{\rm e}^{\beta(E_0+\mu)}}{1 + {\rm e}^{\beta(E_0+\mu)}}~=~{\rm const}. \label{eq:proofgrandcanores}
\end{eqnarray}
Therefore, the agent distribution corresponds to the degree distribution scaled by a constant coefficient $D$; they agree with each other, and the $R^2$ value becomes one after normalization. 
As described later, characteristic quantities, such as the hunting and imbalance rates, show slight periodic undulations in the time direction; \HLL{however, these undulations are typically less than one percent and are negligible by assuming an equilibrium state.} 

\HLLL{The advantage of proving the agreement between the agent and degree distributions using the grand canonical ensemble in statistical physics is that Eq.~(\ref{eq:proofgrandcanores}) can be applied in infinite networks because our description requires only the assumption of equilibrium states for each node and an identical activity rate for all nodes. In contrast, the previous explanation assumes a finite network with a stationary density distribution to define the transition probability matrix~\cite{MASUDA20171}. Infinite networks differ significantly from finite networks in the random walk behaviors of the agents, such as that agent recursion is not guaranteed. Our description shows that in real-world systems, even networks that can be considered infinite networks, such as neural networks, supercrystals~\cite{doi:10.1080/10610270108027472, doi:10.1021/cg034199d, doi:10.1021/acs.inorgchem.6b01199}, and biomolecular networks~\cite{doi:10.1021/nn502149u}, are guaranteed to match these two distributions as long as the above conditions are satisfied. Furthermore, because the activity rates among the nodes are the same, this model is considered suitable for homogeneous networks. This is the first study to prove that these two distributions coincide in infinite networks using the grand canonical ensemble of statistical physics.} Accordingly, we demonstrated that the degree distribution precisely determines the agent distribution when the agents perform a fully random walk. 

Two significant findings were obtained. First, the accuracy of the agreement between the agent and degree distributions depends on the ratio of the number of agents to the number of nodes $N_{a}/N_{n}$. Figure~\ref{fig:Figure2}(D) shows the dependence of $R^2$ on $N_{a}/N_{n}$ for the two cases, $N_n = 30$ and $N_n = 300$. $R^2$ values are confirmed to converge to one when $N_{a}/N_{n}$ is approximately greater than 60 in both cases. That is, the agent and degree distributions become congruent although the number of nodes increases while the number of agents is sufficient, such as $N_{a}/N_{n} > 60$. Second, the agents' congestion-avoiding behavior alters the agent distribution from one with a fully random walk in the steady state. The red bars in Rule (b) of Fig.~\ref{fig:Figure2}(E) show that the $R^2$ value decreases compared with Rule (a). Although Rule (b) differs only from Rule (a) in that each agent excludes the most congested adjacent node as a destination before it starts walking randomly, this difference alters the overall agent distribution. A similar observation is true and more distinctive in Rule (c), where agents always select the sparsest adjacent node. The $R^2$ value then further decreases compared with Rule (b), as shown by the white bars in Fig.~\ref{fig:Figure2}(E).

\begin{figure*}[t]
\vspace{-1.0cm}
\includegraphics[width=0.96\textwidth, clip, bb= 0 0 3450 4550]{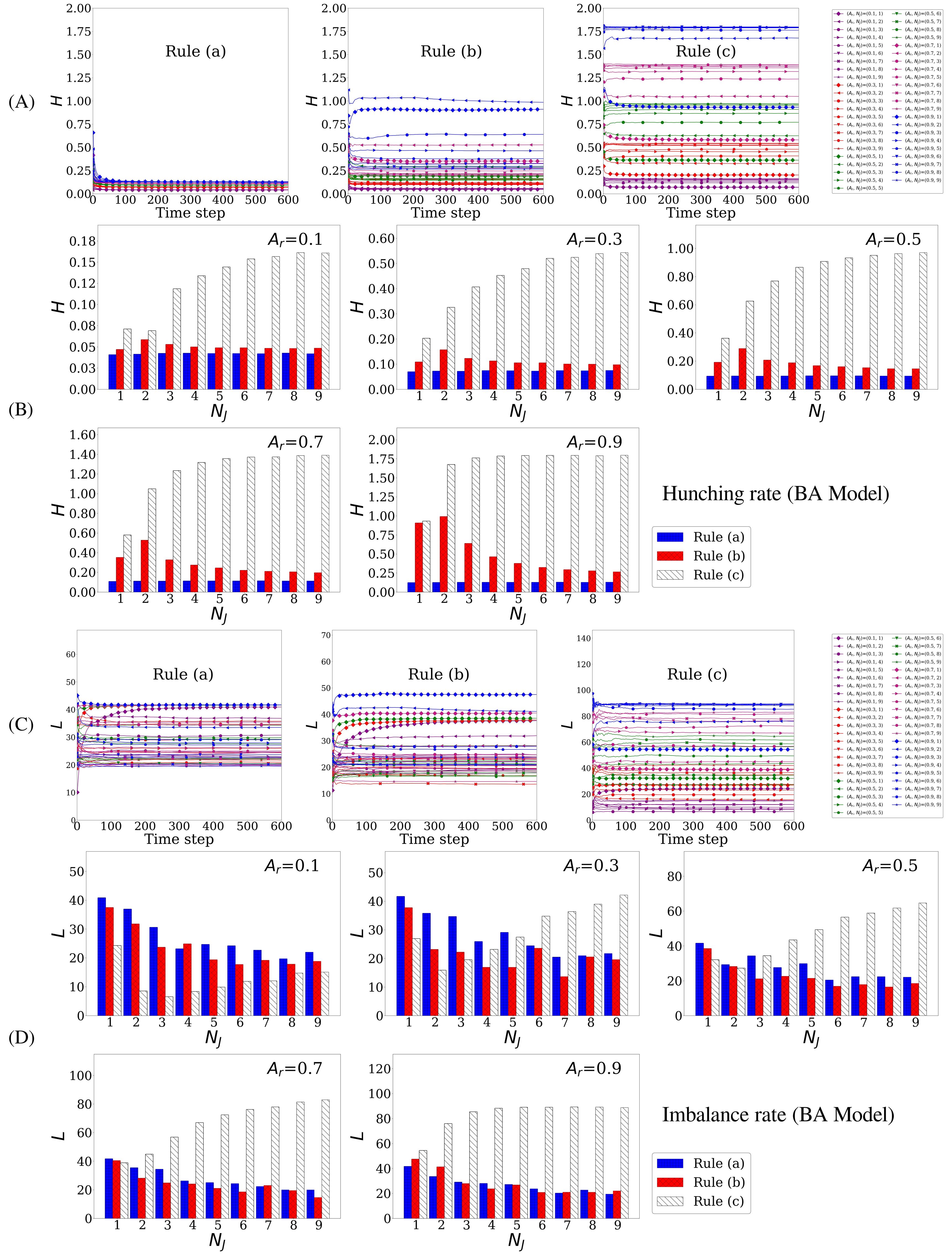}
\caption{(A) Time-dependence of the hunting rate $H$ variation on BA network models for all the measured cases with three different rules: (a), (b), and (c). (B) Comparison of $H$ for five different activity rates $A_r$ between 0.1 and 0.9; each subfigure shows a histogram of $H$ for nine different characteristic network parameters $N_J$ with the three different rules: (a), (b), and (c). (C) Time-dependence of the imbalance rate $L$ variation for all the measured cases in the three different rules: (a), (b), and (c). (D) Comparison of $L$ for five different activity rates $A_r$ between 0.1 and 0.9; each subfigure shows a histograms of $L$ for nine different network parameters $N_J$ with the three rules: (a), (b), and (c).}
\label{fig:Figure4}
\end{figure*}

\begin{figure*}[t]
\vspace{-1.0cm}
\includegraphics[width=0.96\textwidth, clip, bb= 0 0 3450 4550]{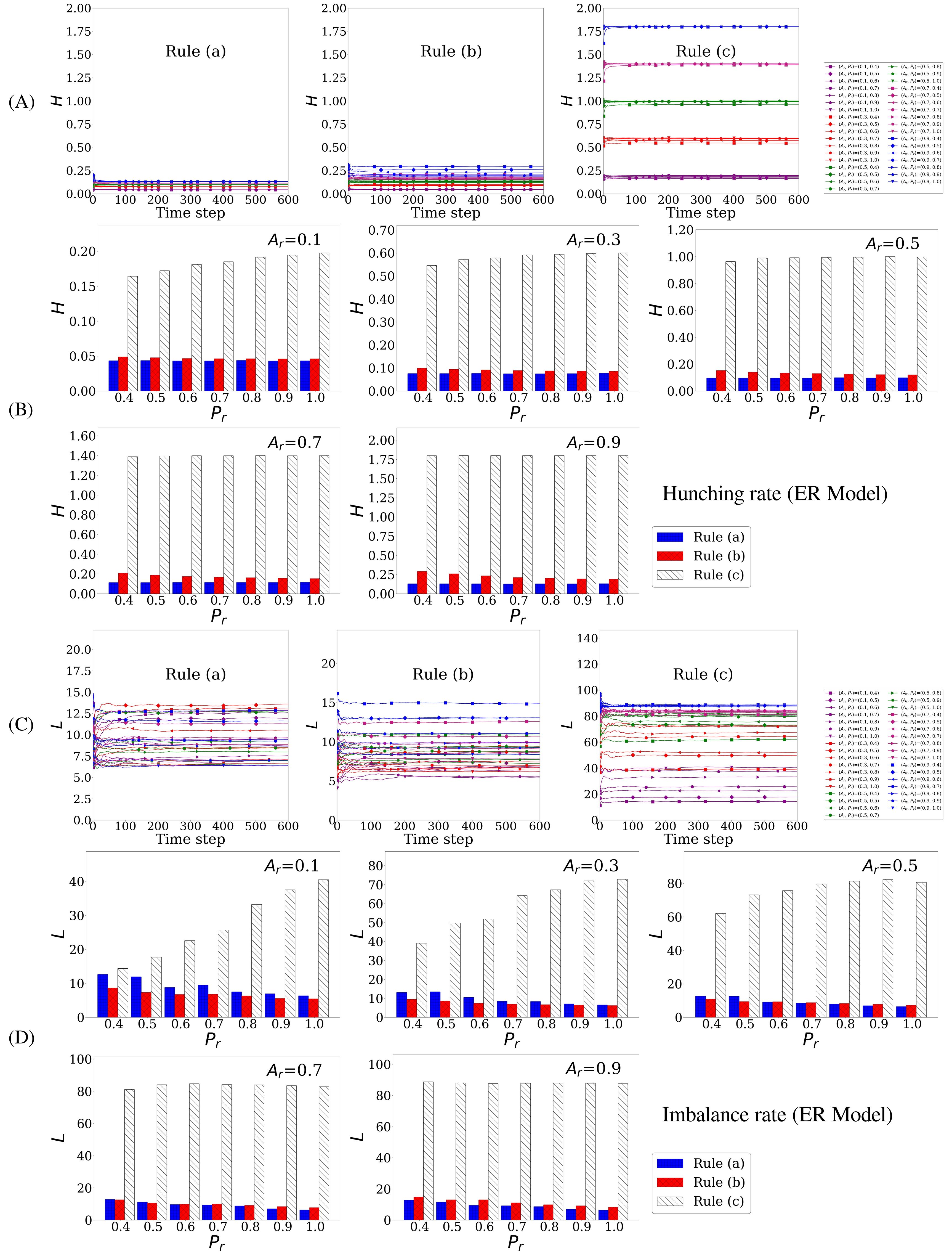}
\caption{(A) Time-dependence of the hunting rate $H$ variation on ER network models for all the measured cases with three different rules: (a), (b), and (c). (B) Comparison of $H$ for five different activity rates $A_r$ between 0.1 and 0.9; each subfigure shows a histogram of $H$ for seven different characteristic network parameters $P_r$ with the three different rules: (a), (b), and (c). (C) Time-dependence of the imbalance rate $L$ variation for all the measured cases with the three different rules: (a), (b), and (c). (D) Comparison of $L$ for five different activity rates $A_r$ between 0.1 and 0.9; each subfigure shows a histograms of $L$ for seven different network parameters $P_r$ with the three rules: (a), (b), and (c).}
\label{fig:Figure5}
\end{figure*}

\begin{figure*}[t]
\vspace{-1.0cm}
\includegraphics[width=0.96\textwidth, clip, bb= 0 0 3450 4550]{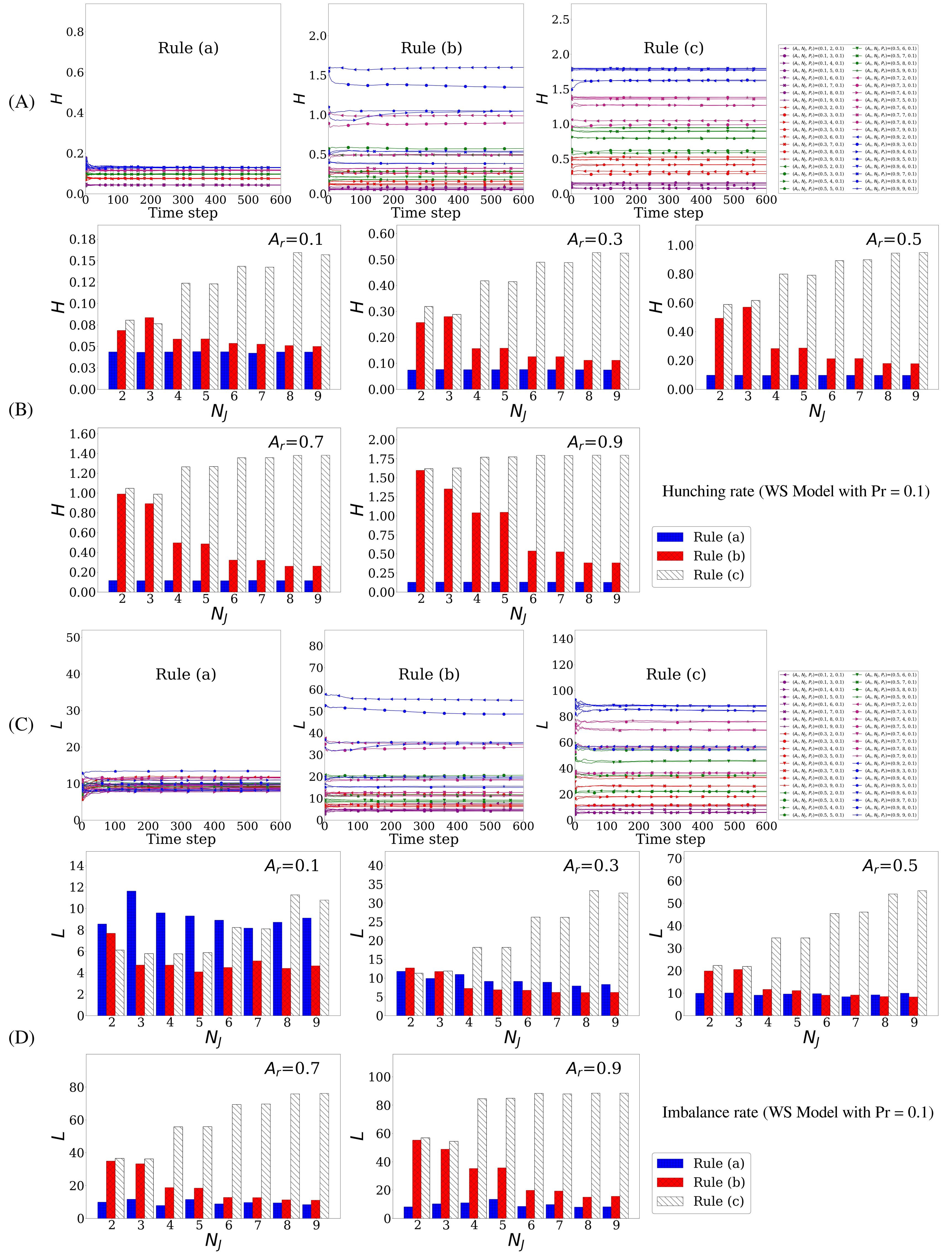}
\caption{(A) Time-dependence of the hunting rate $H$ variation on WS network models for all the measured cases with three different rules: (a), (b), and (c) when $P_r$ = 0.1. (B) Comparison of $H$ for five different activity rates $A_r$ between 0.1 and 0.9; each subfigure shows a histogram of $H$ for eight different characteristic network parameters $N_J$ with the three different rules: (a), (b), and (c). (C) Time-dependence of the imbalance rate $L$ variation for all the measured cases with the three different rules: (a), (b), and (c). (D) Comparison of $L$ for five different activity rates $A_r$ between 0.1 and 0.9; each subfigure shows a histogram of $L$ in eight different network parameters $N_J$ for Rules (a), (b), and (c).}
\label{fig:Figure6}
\end{figure*}

\begin{figure*}[t]
\vspace{-1.0cm}
\includegraphics[width=0.96\textwidth, clip, bb= 0 0 3450 4550]{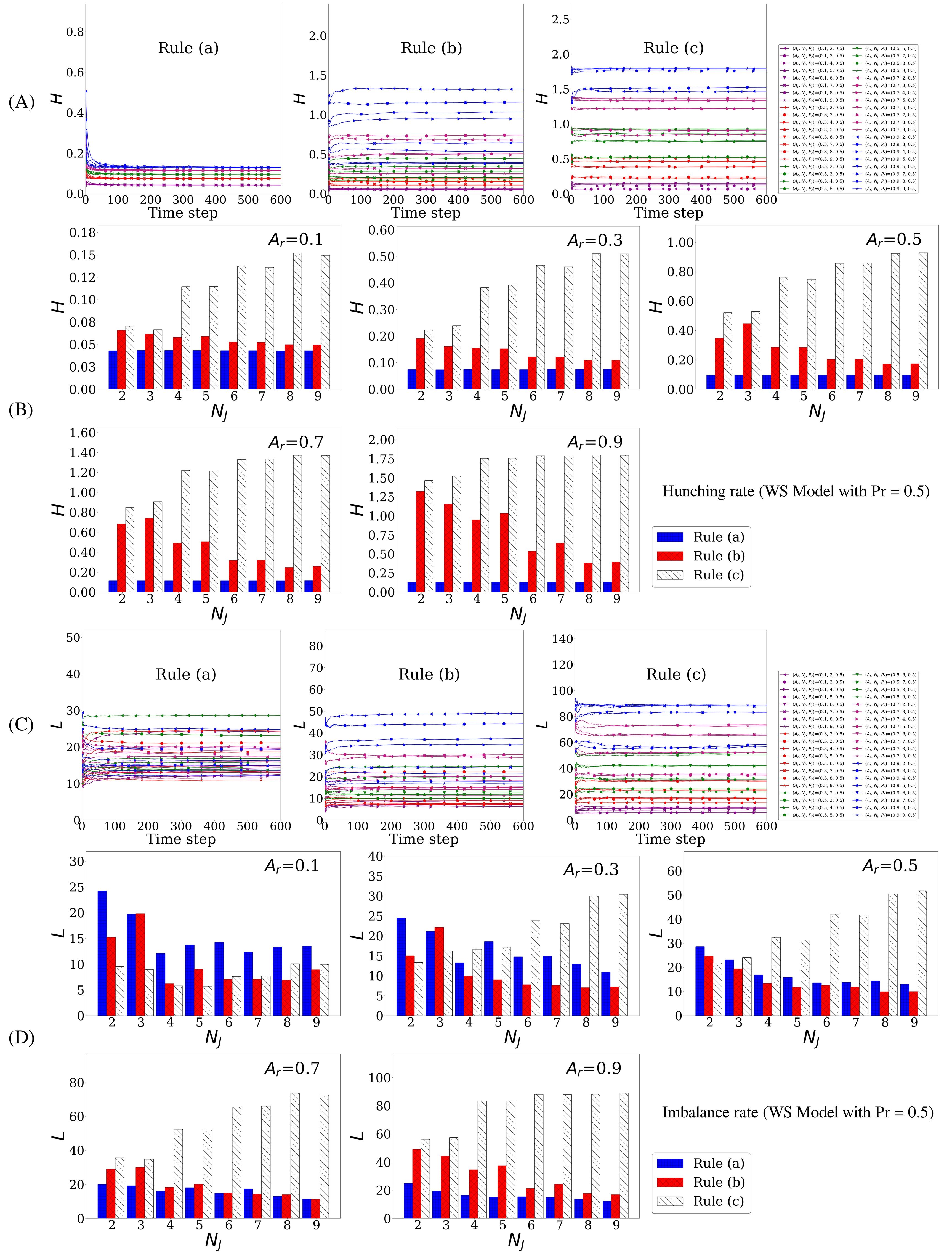}
\caption{(A) Time-dependence of the hunting rate $H$ variation on WS network models for all the measured cases with three different rules: (a), (b), and (c) when $P_r$ = 0.5. (B) Comparison of $H$ for five different activity rates $A_r$ between 0.1 and 0.9; each subfigure shows a histogram of $H$ for eight different characteristic network parameters $N_J$ with the three different rules: (a), (b), and (c). (C) Time-dependence of the imbalance rate $L$ variation for all the measured cases with the three different rules: (a), (b), and (c). (D) Comparison of $L$ for five different activity rates $A_r$ between 0.1 and 0.9; each subfigure shows a histogram of $L$ for eight different network parameters $N_J$ for Rules (a), (b), and (c)}.
\label{fig:Figure7}
\end{figure*}

\begin{figure*}[t]
\vspace{-1.0cm}
\includegraphics[width=0.96\textwidth, clip, bb= 0 0 3450 4550]{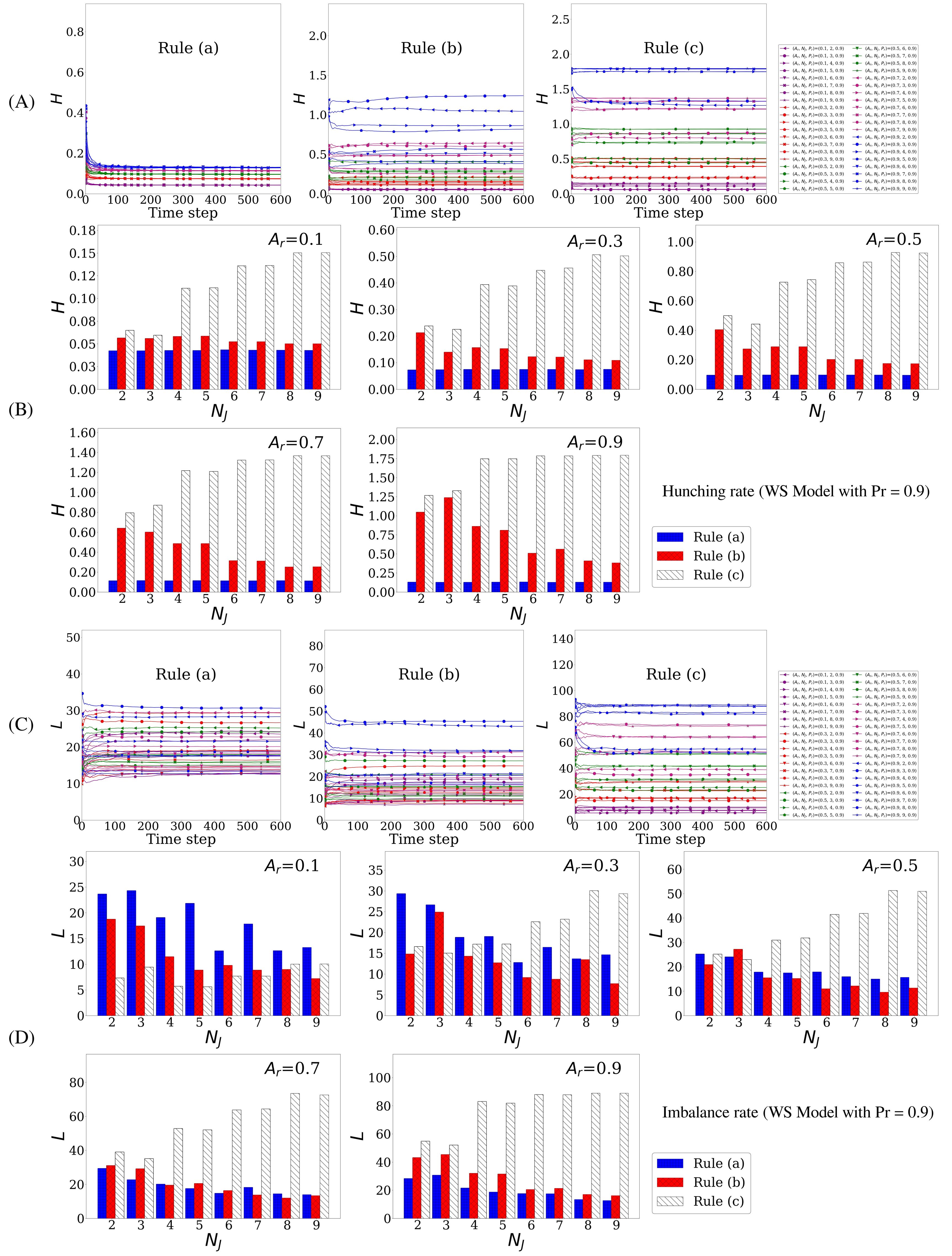}
\caption{(A) Time-dependence of the hunting rate $H$ variation on WS network models for all the measured cases with three different rules: (a), (b), and (c) when $P_r$ = 0.9. (B) Comparison of $H$ for five different activity rates $A_r$ between 0.1 and 0.9; each subfigure shows a histogram of $H$ for eight different characteristic network parameters $N_J$ with the three different rules: (a), (b), and (c). (C) Time-dependence of the imbalance rate $L$ variation for all the measured cases with the three different rules: (a), (b), and (c). (D) Comparison of $L$ for five different activity rates $A_r$ between 0.1 and 0.9; each subfigure shows a histogram of $L$ for eight different network parameters $N_J$ for Rules (a), (b), and (c)}.
\label{fig:Figure8}
\end{figure*}

Next, we measured the hunting rate $H$ and imbalance rate $L$ averaged for the elapsed time step by varying the $N_J$ values between 1 and 9 for the BA model, $P_r$ values ranging from 0.4 to 1.0 for the ER model, and $N_J$ values between 2 and 9, with $P_r$ set to 0.1, 0.5, or 0.9 for the WS model. We collected 200 simulation results for each rule and compared Rules (a)--(c) using 600 different datasets. We report the effect of rule differences on the agent dynamics in complex networks extracted from these datasets. First, our findings show that for a network with a high degree caused by the high values of $N_J$ or $P_r$, which are approximately 8--9 for $N_J$ and 0.8--1.0 for $P_r$, Rule (c), with which agents always select the sparsest adjacent node, drastically \HLL{increases} the hunting rate $H$ and imbalance rate $L$ compared with the random-walk scenarios, as in (a) and (b). We can \HLL{observe} such \HLL{surges in $H$ and $L$} \HLLLL{in Rule (c)} for all three network types in the high degree ranges, as shown in all subfigures of Figs.~\ref{fig:Figure4}--\ref{fig:Figure8}. \HLL{The exceptional cases} are the imbalance rate $L$ of the BA and WS models at $A_{r} = 0.1$ in Fig.~\ref{fig:Figure4} and \ref{fig:Figure6}--\ref{fig:Figure8}, for which the activity rates were sufficiently low. 
\HLLLL{Second, the aforementioned surges in $H$ and $L$} with Rule (c) \HLLLL{were} most distinctive with the ER model, as shown in Fig.~\ref{fig:Figure5}, for all measured cases. \HLLLL{In contrast}, \HLLLL{they were} not necessarily evident in the BA and WS models when the network has a low degree owing to small values of $N_J$ and $P_r$ such as $N_{J} = 2$ and $P_{r} = 0.4$. From a collective dynamics viewpoint, the results can be \HLLLL{interpreted} as follows: when each individual agent intentionally selects the sparsest adjacent node as their destination, both the hunting and imbalance rates drastically increase regardless of the network type, as long as the network has a high degree and activity rate, compared with the cases in which agents walk randomly. However, this phenomenon is not necessarily observed when the network has a low degree or activity rate.

Moreover, compared with random walk-based Rules (a) and (b), we found that Rule (b), with which agents exclude the most congested adjacent nodes as a destination, increases the hunting rate $H$ and concurrently decreases the imbalance rate $L$ compared with Rule (a) when the activity rate $A_{r}$ of the agents is small, such as $A_{r} = 0.1$. This trade-off relationship is valid for all network types when $A_r = 0.1$ for almost all $N_{J}$ and $P_{r}$ parameter ranges. For example, we compare the cases at $A_r=0.1$ for (B) and (D) in Fig.~\ref{fig:Figure5}. We can confirm that the red bar for Rule (b) is greater than the blue bar for Rule (a) for the hunting rates in (B), whereas the relations are inversed for the imbalance rates of the corresponding cases in (D). Similar observations are made for the other network models when $A_r = 0.1$ and $A_r = 0.3$, as shown in Fig.~\ref{fig:Figure4}--\ref{fig:Figure8}. Conversely, Rule (b) increases both $H$ and $L$ compared with Rule (a) when $A_r$ is sufficiently large, such as $A_{r} = 0.9$. From a collective dynamics perspective, these observations can be explained as follows. In principle, avoiding the most congested nodes effectively mitigates the imbalance in agent distribution as the agents are distributed. However, agents move to less-congested nodes simultaneously. Thus, the congestion avoidance behavior in Rule (b) reduces the imbalance rate and increases the hunting rate. This is why the hunting and imbalance rates exhibit a trade-off relationship when the activity is low. However, when group activity is too high, the hunting rate increases, and the magnitude of the dispersion effect from avoiding congestion remains at the same level. In contrast, most agents at the currently congested node move to another node the next time owing to high activity; however, the agents do not select the currently congested nodes as their destinations because they determine them based on the congestion information at the current time. Consequently, the congested node becomes one of the sparsest adjacent nodes the next time. This mismatch between destinations increases the imbalance rate. Accordingly, congestion avoidance behavior can significantly increase the hunting and imbalance rates. 
These observations are distinctive for the ER and WS models because they have a relatively uniform degree distribution, which are not necessarily true for the BA model because of its localized distribution. In the BA model, the agent distribution is localized owing to the localized degree distribution. Nodes with higher degrees tend to have a large number of agents, even with Rule (b), as shown in (B) in Fig.~\ref{fig:Figure2}. Therefore, nodes with higher degrees can generate more significant destination mismatches, leading to high imbalance rates. Because the agent distribution depends on the degree distribution, the obtained imbalance rate differs depending on the parameter $N_J$, as shown in the subfigure for $A_{r} = 0.9$ in (D) of Fig.~\ref{fig:Figure4}. 

\HLLL{In summary, we made the following two observations regarding decision-making strategies. First, the agents' decision to always select the sparsest adjacent node among neighboring nodes makes both the hunting and imbalance rates significantly higher compared with the random-walk scenarios, particularly when the network has a high degree and activity rate; this is observed in all three network models. Second, the agents' decision to exclude the most congested adjacent node as a destination increases the hunting rate while decreasing the imbalance rate when agent activity is low. Conversely, when agent activity is high, their decisions increase both the hunting and imbalance rates. The second observation is distinctive for the ER and WS models because they have relatively uniform degree distribution, but this is not necessarily valid for the BA model because of the localized agent distribution. Figure~\ref{fig:Figure3} summarizes these observations for the ER model. Here, parameters $H$ and $L$ are respectively scaled by the larger of the measured values in each comparison.} \HLLLL{It was confirmed from Figs.~\ref{fig:Figure3}-\ref{fig:Figure8} that relationship (1) in Fig.~\ref{fig:Figure3} was true for all three network models. Moreover, relationships (2) and (3) in Fig.~\ref{fig:Figure3} were true for the ER and WS models; however, they were not necessarily observed in the BA model.} 

As a further step in our research, we performed a Fourier analysis of the time-dependent variation in the hunting and imbalance rates. In the previous analysis, the average values of Eqs. (\ref{eq:defparaH}) and (\ref{eq:defparaL}), over time. However, in real-world systems, we are interested in the temporal variation of these characteristic quantities, in addition to the mean, because short-term fluctuations in these physical quantities can cause state transitions, such as crowd avalanches and agent collisions. We obtained the autocorrelation functions of the time series data in Eqs. (\ref{eq:defparaH}) and (\ref{eq:defparaL}) by multiplying the data using the Fourier transform and inverse Fourier transform of the power spectrum. Figure~\ref{fig:Figure9} displays all the results of the obtained autocorrelation functions for 200-time series data, where $S_h$ and $S_l$ denote the values of the autocorrections of $H$ and $L$, respectively. The vertical axes of the respective subfigures in the left columns show the results of $S_h$, and those of the subfigures in the right columns display those of $S_l$. \HLL{The horizontal axis in each subfigure represents the scaled time $T_s$, which is obtained by dividing the number of time steps by the total number of time steps $N_t$. Each data is shifted by $N_t/2$ to center the time base point. Each subfigure is enlarged such that the midpoint of the maximum and minimum values is located at the center of the subfigure in the vertical direction, and the maximum and minimum values fit into the figure.} The common denominator for all the data is that the autocorrelation function shows periodic undulations over multiple time steps for several time periods; \HLL{however, these undulations are typically less than one percent.} The patterns of the periodic undulation were case-by-case. Some results are expanded and highlighted, and other results are shown as thumbnails \HLLL{for reference}.
\HLLL{Accordingly}, we confirmed that the hunting and imbalance rates exhibit periodic undulations over time.

\section{Discussion}\label{sec:discuss}
First, we summarize our findings. Our \HLLLL{simulation} study supports the following findings on the effect of the topological network structure and the three different effects of agents' destination selection on the collective dynamics of agents in three complex networks: the BA, ER, and WS models.

\begin{itemize}
\item [(A)] The degree distribution, that is, the topological structure of a network, precisely determines the shape of the agent distribution among nodes in the steady state when agents perform a full random walk. 
\HLL{We demonstrated that the theory of a grand canonical ensemble in statistical physics describes this well.}
In addition, their decisions on selecting destinations alter the overall agent distribution.

\item [(B)] The agents' decision to always select the sparsest adjacent node among neighboring nodes makes both the hunting and imbalance rates significantly high compared with the random-walk scenarios, particularly when the network has a high degree and activity rate; this is observed in all three network models.

\item [(C)] The agents' decision to exclude the most congested adjacent node as a destination increases the hunting rate while decreasing the imbalance rate when agent activity is low. Conversely, when agent activity is high, their decisions increase both hunting and imbalance rates. This is observed for the ER and WS models because they have relatively uniform degree distribution, and it is not necessarily valid for the BA model because of the localized agent distribution.

\item [(D)] 
The hunting and imbalance rates show \HLL{slight} periodic undulations with respect to time. 
\end{itemize}

\HLLL{Regarding (A), many studies would have empirically acknowledged in their respective fields that the degree distribution of a network can contribute to the formation of an agent distribution, even if the network is an infinite network such as a neural, supercrystal, or biomolecular network. However, mathematically proving the agreement between agent and degree distributions is difficult because we cannot define a stationary probability density distribution for infinite networks owing to its openness; it follows that a transition probability matrix to the network cannot be defined. This study is the first to quantitatively ensure that the degree distribution determines the shape of the agent distribution for general cases, including infinite networks, using a theoretical model of a grand canonical ensemble in statistical physics.}
\HLJPCOMPX{In addition, Eq.~(\ref{eq:proofgrandcanores}) assumes a grand canonical ensemble in each node and the condition that the activity rate is the same for all nodes, and the latter allows us to derive Eq.~(\ref{eq:proofgrandcanores}) from Eq.~(\ref{eq:grandcanonanop}). Here, we can rephrase these conditions as each node reaching an equilibrium state where the amount of inflow/outflow is at least quasi-statically equal in each node and its magnitude is equal everywhere. They can be satisfied as long as there are fully random walks in complex networks that have no isolated nodes. A good example is the PageRank~\cite{Chung2010, LAI20102443} distribution over web pages, which is known to be a fully random walk problem. However, this proof does not hold if there are differences in the scale of the activity rate in the respective nodes or if the independence of the agents is broken by behaviors such as congestion avoidance. Nevertheless, the statistical mechanical explanation of the correspondence between agent and degree distributions discussed here can be applied to real cases, such as epidemic metapopulation models, which in certain cases assume a fully random walk in a network connecting subpopulations~\cite{NAGATANI201866}.}

The application of findings (A)--(C) to real-world collective dynamics can be described as follows. First, (A) allows us to estimate the agent distribution from the degree distribution, which is the topological structure of the network, in situation in which agents can be regarded as performing fully random walks, such as in crowds, metapopulations, and the dynamics of neurotransmitters in the brain. Once we identify the hubs of the nodes from the degree distribution, we can predict the congested areas of crowds because of the agreement between the agent and degree distribution, which may help prevent crowd incidents. (A) also helps us construct facilities where people congregate. By referring to (A), we can design these facilities to avoid the formation of dense crowds by eliminating the hubs in the network. (A) is also applicable in systems that guide vehicles, that is, for agents in highway parking and air and ground transportation systems with multiple lanes for arriving and departing aircraft. If the degree of the distribution of parking and ground traffic conduits is known, we can predict the locations where congestion may occur. 

(B) and (C) significantly contribute to accident prevention. Our results suggest that agents avoiding the most congested adjacent nodes concurrently increases the hunting rate and decreases the imbalance rate when the crowd is inactive. Therefore, if crowd managers need to prioritize dispersing people in situations where people are not very active or stuck, instructing agents to avoid the most congested sites is effective in distributing people while admitting an increase in the movement of agents per unit time. However, we revealed that intentionally conducting avoidance behaviors increases both the hunting and imbalance rates when the crowding activities are high; this instruction should be averted in such cases. In addition, we emphasize that the agents' decision to select the sparsest adjacent node should be avoided, except for in some sufficiently low-degree networks, based on (B). This selection makes both the hunting and imbalance rates significantly high on a complex network, regardless of the network type. Findings (A)--(C) can be applied to problems involving controlled agents in complex networks. For several ground traffic controls at airports, the airport's geometric structure may create areas where taxis are likely to crowd. According to (A), airport controllers can consciously instruct pilots in advance to avoid concentrating on taxing aircraft. At this time, directing aircraft to the sparsest adjacent spots may significantly further imbalance the airport spot occupancy and number of aircraft movements per unit time owing to (B). Furthermore, we can understand from (C) that weighing the advantages of directing aircraft to move to other locations would be beneficial in reducing the imbalance rate against the disadvantages of fuel consumption and additional costs involved in moving aircraft in a short period.

\HLJPCOMPX{Nevertheless, we must discuss the difference between the real cases and the cases with Rules (a)--(c) when applying our results to real cases. Recall an important result. In Rule (b), avoiding the most congested nodes effectively mitigates the imbalance in agent distribution as agents are distributed among remaining nodes. However, the number of agents in less congested nodes simultaneously increases. Therefore, the congestion avoidance behavior in Rule (b) reduces the imbalance rate and increases the hunting rate when the activity rate is low; in other words, we can see a trade-off relationship between Rules (a) and (b), as shown in Fig.~\ref{fig:Figure3}, when the activity is low. However, as the group activity becomes high, the hunting rate increases and the magnitude of the dispersal effect of avoiding congestion remains at the same degree of avoiding the most congested node. Consequently, both the hunting rate and the imbalance rate increase under Rule (b) compared to Rule (a) when the activity rate is high. In real cases, however, agents may still choose the most congested node as their destination if the node is still attractive, even considering its congestion. This is exemplified by the situation where agents can get some incentive (e.g., food or services limited at the node). Thus, if practical conditions exist such that the node has a priority, an external influence that forces agents to the node, or a high popularity among agents, the results will differ from the case of Rule (b). We should consider these points when extending Rule (b) to real cases, where the complexity of decision-making holds. 

Recall another important finding about agent dynamics with Rule (b) when the activity rate is high. In this case, agents at the currently most congested node move to another node at the next time step. However, agents at the other nodes do not choose the currently most congested node as their destination to avoid congestion; this decision-making is based on the congestion information at the current time. As a result, the most congested node becomes one of the most sparse neighboring nodes the next time. In short, destination mismatches increase the congestion rate. Accordingly, Rule (b) significantly increases hunting and imbalance rates compared to Rule (a). These observations were particularly evident for the ER and WS models, which have relatively uniform degree distributions. Importantly, the aforementioned destination mismatches occur because agents simultaneously decide their destinations based on the current congestion information. However, in real cases, agents can make more comprehensive decisions based on current and past time congestion information, namely, predicting whether the congestion at the most congested node is transient or persistent based on experience. In short, agents can choose the currently most congested node as their destination when they know that the node will become emptier at the next time. This strategic decision with a higher level of thinking can prevent the simultaneous migration of large agents, which mitigates the hunting rate and the imbalance rate. Therefore, we must consider this point when extending the obtained fact that destination mismatches increase the hunting and imbalance rates to real cases.}

\begin{figure*}[t]
\vspace{-1.3cm}
\includegraphics[width=0.99\textwidth, clip, bb= 0 0 3500 4250]{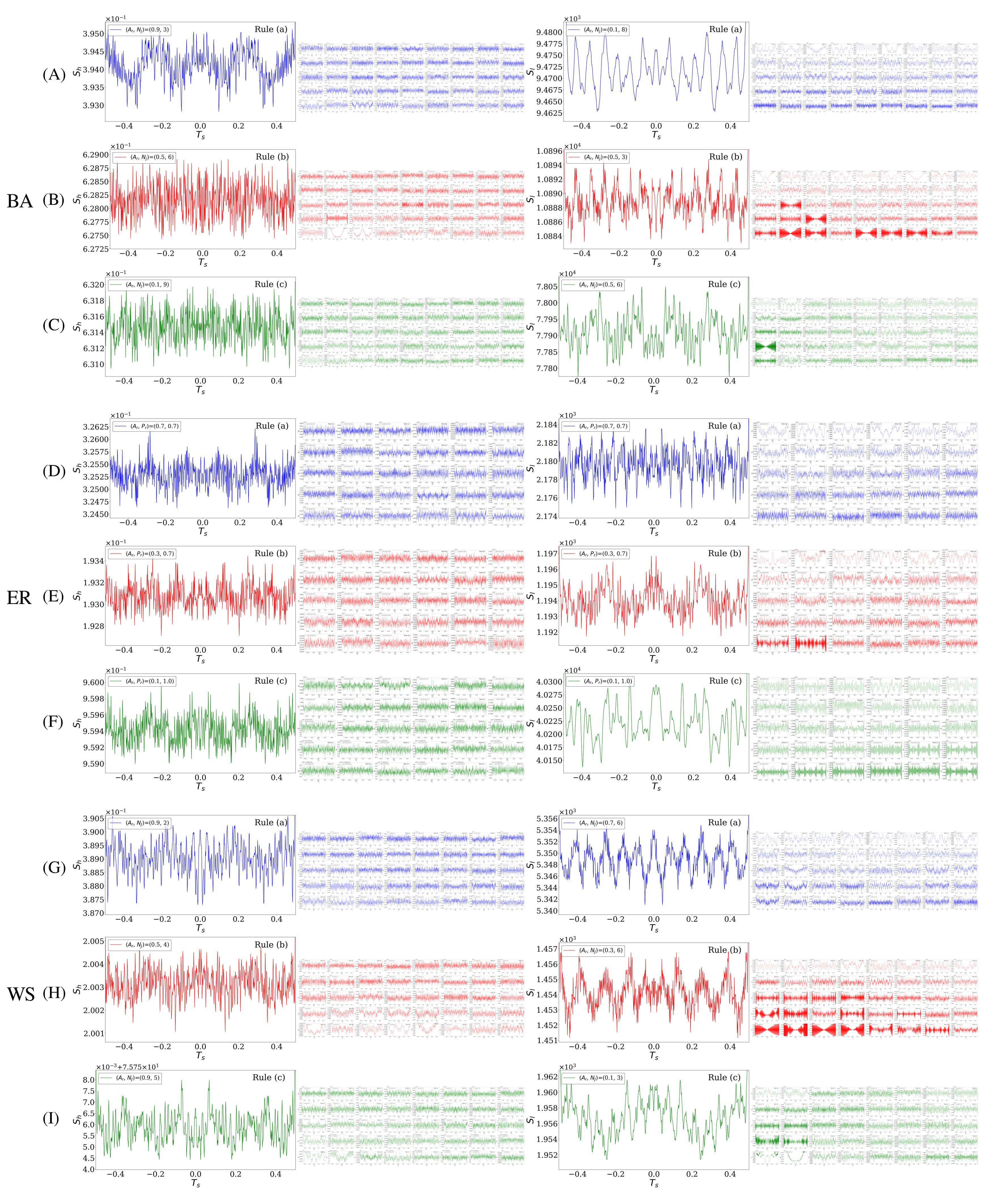}
\caption{Variation of the autocorrelation of hunting rate $H$ (left) and imbalance rate $L$ (right) on different complex networks. (A)--(C) show the results of the BA models, where (A), (B), and (C) correspond to the results of Rules (a), (b), and (c), respectively. Similarly, (D)--(F) show the results of the ER models, where (D), (E), and (F) correspond to the results of Rules (a), (b), and (c), respectively. (G)--(I) show the results of the WS models with $P_r$=0.5, where (G), (H), and (I) correspond to the results of Rules (a), (b), and (c), respectively.}
\label{fig:Figure9}
\end{figure*}

Finally, (D) may contribute to the field of neuroscience more than engineering. The quasi-periodic patterns of neural activity in the brain contribute to typical brain functions ~\cite{ABBAS2019193, Belloy2018}. In addition, this function has been reported to be reduced in patients with ADHD \cite{ABBAS2019101653}. Here, Fig.~\ref{fig:Figure9} reported that the phase of the periodic undulation of network characteristics differs depending on the network type and moving rules of the agents. If we consider the brain as a complex network and the agents as signals produced by neuronal interactions, the periodic undulation of network properties reported in (D) may represent the simplest form of periodic patterns of neural activity. Therefore, the correspondence between our results and those of real-world systems must be further investigated in detail. Nonetheless, in this manner, we can say that our findings are helpful for the agent mobility problem in complex networks.
\HLJPCOMPX{As shown in (A) and (C) in Figs.~\ref{fig:Figure4}-\ref{fig:Figure8}, the averaged imbalance rate and the hunting rate reach a steady state within the time and space scale. However, there was a possibility that we missed the slight periodic undulations in the original time series, which is negligible after the averaging operation. In terms of risk management in the crowd, it is more important to detect the variation of $H$ and $L$ in local time and space than the variation of their averages. This is because the abrupt increase in $H$ in a small period of time indicates the occurrence of rapid mass migration, which could cause accidents such as collisions among agents in real cases. If we find the existence of such a slight periodic undulation and the correlation between its periodic behavior and the abrupt variation of $H$ or $L$ in a short time, we can use it for crowd control by predicting the occurrence of accidents in advance. However, to the best of our knowledge, no study has investigated the periodic undulations of the feature value of multiple agents in complex networks using CA simulations. Therefore, as a preliminary step for future studies, we focused on revealing the existence of such periodic undulations in the respective rules. As shown in Fig.~\ref{fig:Figure9}, this study confirmed the existence of slight periodic undulations to a certain extent. Recall that we focused on the Markovian process (i.e., the state at $n+1$ time step determined only from $n$ time steps) for all analyses in this research. In this respect, a similar discussion on non-Markovian agent networks~\cite{Kyriakis2020, 8815074} may allow us to discuss the system closer to real cases; this should be done in future work.}

\HLJPCOMPX{As a prelude to future studies, we discuss the resilience of the system under our scenarios. We performed agent simulations in which agents follow Rules (a)--(c) on BA, ER, and WS network models with a functionality of disconnecting nodes in a certain time period during the simulation; the results are shown in Fig.10. Note that we set the parameters $(N_n, N_a, N_t)$ to $(30, 2000, 600)$, similar to the other cases. In the simulations, we randomly selected nodes with a probability of 50 $\%$ at the time of 300 time steps, then isolated the selected nodes by forbidding agents to travel to them during the time from 300 to 400 time steps. We then reconnected these isolated nodes to networks by allowing agents to access them when the simulation reached 400 time steps. We measured the time variation of the mean value of the hunting and imbalance rates in all measurable cases for BA and ER, and $P_r$ = 0.5 for WS, similar to Figs.~\ref{fig:Figure4}, \ref{fig:Figure5} and \ref{fig:Figure7}; therefore, the measured cases of H and L for each network model correspond to (A) and (C) in Figs.~\ref{fig:Figure4}, \ref{fig:Figure5} and \ref{fig:Figure7}, respectively. 

The following results were obtained. First, it was found that the hunting rate $H$ tends to decrease and the imbalance rate $L$ tends to increase as soon as the isolation of a node starts, regardless of the network type. When the network is restored, both the hunting rate $H$ and the imbalance rate $L$ are found to recover gradually (in some cases not completely within the measurement time). The reason why $H$ decreases and $L$ increases with node isolation can be explained as follows. When a node is isolated, there are fewer candidates as destinations for agents. The imbalance rate represents the average deviation between the number of agents per node when agents are evenly distributed and the actual number of agents in each node. However, as the number of agents at the isolated node is forced to always be zero, these deviations increase, and the imbalance rate increases compared to before the isolation operation. Meanwhile, the hunting rate represents the average change from the previous time in the number of agents for each node. Since node isolation reduces the number of candidate nodes as destinations for agents to visit, agents are forced to travel to remaining nodes other than the isolated nodes. Because the same number of agents is spread over fewer nodes, there will always be a certain number of agents per node, and the frequency of situations where there are no agents at all at a node will decrease. In addition, since the number of agents in the isolated nodes is zero, their hunting rate is always zero. Thus, the overall hunting rate is reduced by isolation. As mentioned above, the hunting rate is a measure of the average amount of migration per unit of time, so it is desirable for the hunting rate to be small in order to reduce the risk of pedestrian collisions and other occurrences in real systems. In a particular transport network, if the hunting rate is large, it may be possible to reduce it by daring to shut down some nodes. Of course, this would need to be verified in more realistic scenarios and networks.}

\section{Conclusion}
We examined the agent dynamics in three complex networks: the BA, ER, and WS models under the following rules: agents (a) randomly select a destination among adjacent nodes, (b) exclude the most congested adjacent node as a potential destination and randomly select a destination among the remaining nodes, or (c) select the sparsest adjacent node as a destination. We measured two significant characteristics in the wide parameter ranges of the network characteristic parameters: the hunting rate, which is the change rate of agent amounts in each node per unit time step, and imbalance rate, which is the non-uniformity of agent distribution among nodes. 

Our analysis suggests that the degree distribution, which represents the topological structure of a network, precisely determines the shape of the agent distribution among nodes in steady state when agents perform a full random walk. 
\HLL{The model of a grand canonical ensemble in statistical physics was determined to describe this well.}
However, the decision to select a destination altered the overall agent distribution. Particularly, the agents' decision to always select the sparsest adjacent node among neighboring nodes makes both the hunting and imbalance rates significantly higher compared with random-walk scenarios, particularly when the network has a high degree and activity rate, which was observed in all three network models. In contrast, the agents' decision to exclude the most congested adjacent node as a destination increases the hunting rate while decreasing the imbalance rate when agent activity is low. Conversely, when agent activity is high, their decisions increase both the hunting and imbalance rates. This was observed for the ER and WS models because they had a relatively uniform degree distribution, and was not necessarily valid for the BA model. In addition, the hunting and imbalance rates show periodic undulations with respect to time. 

\HLJPCOMPX{In this paper, we have studied the periodic undulation of two characteristic values of agent behavior, as summarized in Fig.~\ref{fig:Figure9}. Although we have focused on three representative complex networks (BA, ER, and WS models), a similar discussion on the periodic undulation of agent dynamics in differently structured networks, such as multifractal networks~\cite{Yang2021}, would be interesting. This is because such a regular and hierarchical structure could resonate with the periodic undulation of agents and produce unobserved agent behavior in the whole network. Thus, we can say that studying such multifractal networks is a direction for future research. Furthermore, if we are targeting applied real-world cases, it would be better to consider a model where each node or edge has different weights~\cite{Bu2023} rather than focus on the standard model of three complex networks. How different results are obtained compared to the unweighted case treated in this study is also of interest for future studies. To create realistic biological or social networks for our multi-agent analysis, a methodological model that can infer unobserved parts of the network structure would be useful~\cite{Xue2019}. The study of agent dynamics with congestion-avoidance behavior in such a real network should be investigated in the future. In this way, our results have great potential and applicability to various problems involving agent mobility in the complex networks in related fields.}

\section*{Data availability}
All data generated and analyzed during this study are included in this published article.

\section*{Acknowledgment}
This study was supported by the JST-Mirai Program Grant Number JPMJMI20D1, Japan, as well as JSPS KAKENHI Grant Numbers JP21H01570 and  21H01352, and partly supported by MEXT as ``Program for Promoting Research on the Supercomputer Fugaku: Exploring Next Generation Aerospace Mobility and its Extension to Social System via Supercomputer Fugaku (Project ID: hp230198 and JPMXP1020230215)''. 
The authors thank Editage (www.editage.jp) for their English language editing services. 
The authors express special thanks to the administrative staff at Itoh Laboratory and RCAST.
{\bf Author contributions}: S.T., D.Y., E.I., and K.N. designed the study; S.T. conducted the modeling and numerical experiments and data analysis; E.I., D.Y., and K.N. provided funding, project administration, and resources; and S.T. drafted the manuscript. All authors contributed to the feedback and editing of the manuscript. {\bf Competing interests}: The authors declare no conflicts of interest.


\bibliography{reference}

\begin{thebibliography}{10}
\expandafter\ifx\csname url\endcsname\relax
  \def\url#1{\texttt{#1}}\fi
\expandafter\ifx\csname urlprefix\endcsname\relax\def\urlprefix{URL }\fi
\providecommand{\bibinfo}[2]{#2}
\providecommand{\eprint}[2][]{\url{#2}}

\bibitem{doi:10.1080/01441647.2013.848955}
\bibinfo{author}{Lin, J.} \& \bibinfo{author}{Ban, Y.}
\newblock \bibinfo{title}{Complex network topology of transportation systems}.
\newblock \emph{\bibinfo{journal}{Transport Reviews}}
  \textbf{\bibinfo{volume}{33}}, \bibinfo{pages}{658--685}
  (\bibinfo{year}{2013}).
\newblock \urlprefix\url{https://doi.org/10.1080/01441647.2013.848955}.

\bibitem{Xu2011NonliDyn}
\bibinfo{author}{Xu, X.} \emph{et~al.}
\newblock \bibinfo{title}{The chaotic dynamics of the social behavior selection
  networks in crowd simulation}.
\newblock \emph{\bibinfo{journal}{Nonlinear Dynamics}}
  \textbf{\bibinfo{volume}{64}}, \bibinfo{pages}{117--126}
  (\bibinfo{year}{2011}).
\newblock \urlprefix\url{https://doi.org/10.1007/s11071-010-9850-z}.

\bibitem{JABBARPOUR2014303}
\bibinfo{author}{Jabbarpour, M.~R.} \emph{et~al.}
\newblock \bibinfo{title}{Ant-based vehicle congestion avoidance system using
  vehicular networks}.
\newblock \emph{\bibinfo{journal}{Engineering Applications of Artificial
  Intelligence}} \textbf{\bibinfo{volume}{36}}, \bibinfo{pages}{303--319}
  (\bibinfo{year}{2014}).
\newblock
  \urlprefix\url{https://www.sciencedirect.com/science/article/pii/S0952197614001997}.

\bibitem{ARNOTT1991309}
\bibinfo{author}{Arnott, R.}, \bibinfo{author}{{de Palma}, A.} \&
  \bibinfo{author}{Lindsey, R.}
\newblock \bibinfo{title}{Does providing information to drivers reduce traffic
  congestion?}
\newblock \emph{\bibinfo{journal}{Transportation Research Part A: General}}
  \textbf{\bibinfo{volume}{25}}, \bibinfo{pages}{309--318}
  (\bibinfo{year}{1991}).
\newblock
  \urlprefix\url{https://www.sciencedirect.com/science/article/pii/019126079190146H}.

\bibitem{6666580}
\bibinfo{author}{Soylemezgiller, F.}, \bibinfo{author}{Kuscu, M.} \&
  \bibinfo{author}{Kilinc, D.}
\newblock \bibinfo{title}{A traffic congestion avoidance algorithm with dynamic
  road pricing for smart cities}.
\newblock In \emph{\bibinfo{booktitle}{2013 IEEE 24th Annual International
  Symposium on Personal, Indoor, and Mobile Radio Communications (PIMRC)}},
  \bibinfo{pages}{2571--2575} (\bibinfo{year}{2013}).

\bibitem{Tsuzuki2022SciRep}
\bibinfo{author}{Tsuzuki, S.}, \bibinfo{author}{Yanagisawa, D.} \&
  \bibinfo{author}{Nishinari, K.}
\newblock \bibinfo{title}{Effect of congestion avoidance due to congestion
  information provision on optimizing agent dynamics on an endogenous star
  network topology}.
\newblock \emph{\bibinfo{journal}{Scientific Reports}}
  \textbf{\bibinfo{volume}{12}}, \bibinfo{pages}{22159} (\bibinfo{year}{2022}).
\newblock \urlprefix\url{https://doi.org/10.1038/s41598-022-26710-0}.

\bibitem{doi:10.1126/science.1238411}
\bibinfo{author}{Park, H.-J.} \& \bibinfo{author}{Friston, K.}
\newblock \bibinfo{title}{Structural and functional brain networks: From
  connections to cognition}.
\newblock \emph{\bibinfo{journal}{Science}} \textbf{\bibinfo{volume}{342}},
  \bibinfo{pages}{1238411} (\bibinfo{year}{2013}).
\newblock
  \urlprefix\url{https://www.science.org/doi/abs/10.1126/science.1238411}.

\bibitem{Avena-Koenigsberger2018}
\bibinfo{author}{Avena-Koenigsberger, A.}, \bibinfo{author}{Misic, B.} \&
  \bibinfo{author}{Sporns, O.}
\newblock \bibinfo{title}{Communication dynamics in complex brain networks}.
\newblock \emph{\bibinfo{journal}{Nature Reviews Neuroscience}}
  \textbf{\bibinfo{volume}{19}}, \bibinfo{pages}{17--33}
  (\bibinfo{year}{2018}).
\newblock \urlprefix\url{https://doi.org/10.1038/nrn.2017.149}.

\bibitem{ShangYi-Lun_2010}
\bibinfo{author}{Yi-Lun, S.}
\newblock \bibinfo{title}{Multi-agent coordination in directed moving
  neighbourhood random networks}.
\newblock \emph{\bibinfo{journal}{Chinese Physics B}}
  \textbf{\bibinfo{volume}{19}}, \bibinfo{pages}{070201}
  (\bibinfo{year}{2010}).
\newblock \urlprefix\url{https://dx.doi.org/10.1088/1674-1056/19/7/070201}.

\bibitem{10.1063/1.4976959}
\bibinfo{author}{Shang, Y.}
\newblock \bibinfo{title}{{Consensus in averager-copier-voter networks of
  moving dynamical agents}}.
\newblock \emph{\bibinfo{journal}{Chaos: An Interdisciplinary Journal of
  Nonlinear Science}} \textbf{\bibinfo{volume}{27}}, \bibinfo{pages}{023116}
  (\bibinfo{year}{2017}).
\newblock \urlprefix\url{https://doi.org/10.1063/1.4976959}.

\bibitem{liu2015events}
\bibinfo{author}{Liu, C.}, \bibinfo{author}{Zhan, X.-X.},
  \bibinfo{author}{Zhang, Z.-K.}, \bibinfo{author}{Sun, G.-Q.} \&
  \bibinfo{author}{Hui, P.~M.}
\newblock \bibinfo{title}{How events determine spreading patterns: information
  transmission via internal and external influences on social networks}.
\newblock \emph{\bibinfo{journal}{New Journal of Physics}}
  \textbf{\bibinfo{volume}{17}}, \bibinfo{pages}{113045}
  (\bibinfo{year}{2015}).

\bibitem{NIE2021127098}
\bibinfo{author}{Nie, Y.}, \bibinfo{author}{Li, W.}, \bibinfo{author}{Pan, L.},
  \bibinfo{author}{Wang, W.} \& \bibinfo{author}{Lin, T.}
\newblock \bibinfo{title}{Effects of destination selection strategy on
  information spreading}.
\newblock \emph{\bibinfo{journal}{Physics Letters A}}
  \textbf{\bibinfo{volume}{389}}, \bibinfo{pages}{127098}
  (\bibinfo{year}{2021}).
\newblock
  \urlprefix\url{https://www.sciencedirect.com/science/article/pii/S0375960120309658}.

\bibitem{10.1371/journal.pcbi.1006833}
\bibinfo{author}{Avena-Koenigsberger, A.} \emph{et~al.}
\newblock \bibinfo{title}{A spectrum of routing strategies for brain networks}.
\newblock \emph{\bibinfo{journal}{PLOS Computational Biology}}
  \textbf{\bibinfo{volume}{15}}, \bibinfo{pages}{1--24} (\bibinfo{year}{2019}).
\newblock \urlprefix\url{https://doi.org/10.1371/journal.pcbi.1006833}.

\bibitem{10.1063/1.4939837}
\bibinfo{author}{Antonopoulos, C.~G.}
\newblock \bibinfo{title}{{Dynamic range in the C. elegans brain network}}.
\newblock \emph{\bibinfo{journal}{Chaos: An Interdisciplinary Journal of
  Nonlinear Science}} \textbf{\bibinfo{volume}{26}} (\bibinfo{year}{2016}).
\newblock \urlprefix\url{https://doi.org/10.1063/1.4939837}.
\newblock \bibinfo{note}{013102}.

\bibitem{weng2023multiple}
\bibinfo{author}{Weng, T.}, \bibinfo{author}{Chen, X.}, \bibinfo{author}{Ren,
  Z.}, \bibinfo{author}{Xu, J.} \& \bibinfo{author}{Yang, H.}
\newblock \bibinfo{title}{Multiple moving agents on complex networks: From
  intermittent synchronization to complete synchronization}.
\newblock \emph{\bibinfo{journal}{Physica A: Statistical Mechanics and its
  Applications}} \textbf{\bibinfo{volume}{614}}, \bibinfo{pages}{128562}
  (\bibinfo{year}{2023}).

\bibitem{PhysRevE.98.052302}
\bibinfo{author}{Cencetti, G.}, \bibinfo{author}{Battiston, F.},
  \bibinfo{author}{Fanelli, D.} \& \bibinfo{author}{Latora, V.}
\newblock \bibinfo{title}{Reactive random walkers on complex networks}.
\newblock \emph{\bibinfo{journal}{Phys. Rev. E}} \textbf{\bibinfo{volume}{98}},
  \bibinfo{pages}{052302} (\bibinfo{year}{2018}).
\newblock \urlprefix\url{https://link.aps.org/doi/10.1103/PhysRevE.98.052302}.

\bibitem{MASUDA20171}
\bibinfo{author}{Masuda, N.}, \bibinfo{author}{Porter, M.~A.} \&
  \bibinfo{author}{Lambiotte, R.}
\newblock \bibinfo{title}{Random walks and diffusion on networks}.
\newblock \emph{\bibinfo{journal}{Physics Reports}}
  \textbf{\bibinfo{volume}{716-717}}, \bibinfo{pages}{1--58}
  (\bibinfo{year}{2017}).
\newblock
  \urlprefix\url{https://www.sciencedirect.com/science/article/pii/S0370157317302946}.
\newblock \bibinfo{note}{Random walks and diffusion on networks}.

\bibitem{PhysRevE.75.016102}
\bibinfo{author}{Costa, L. d.~F.} \& \bibinfo{author}{Travieso, G.}
\newblock \bibinfo{title}{Exploring complex networks through random walks}.
\newblock \emph{\bibinfo{journal}{Phys. Rev. E}} \textbf{\bibinfo{volume}{75}},
  \bibinfo{pages}{016102} (\bibinfo{year}{2007}).
\newblock \urlprefix\url{https://link.aps.org/doi/10.1103/PhysRevE.75.016102}.

\bibitem{doi:10.1126/science.286.5439.509}
\bibinfo{author}{Barab\'{a}si, A.-L.} \& \bibinfo{author}{Albert, R.}
\newblock \bibinfo{title}{Emergence of scaling in random networks}.
\newblock \emph{\bibinfo{journal}{Science}} \textbf{\bibinfo{volume}{286}},
  \bibinfo{pages}{509--512} (\bibinfo{year}{1999}).
\newblock
  \urlprefix\url{https://www.science.org/doi/abs/10.1126/science.286.5439.509}.

\bibitem{albert2002statistical}
\bibinfo{author}{Albert, R.} \& \bibinfo{author}{Barab\'{a}si, A.-L.}
\newblock \bibinfo{title}{Statistical mechanics of complex networks}.
\newblock \emph{\bibinfo{journal}{Reviews of modern physics}}
  \textbf{\bibinfo{volume}{74}}, \bibinfo{pages}{47} (\bibinfo{year}{2002}).

\bibitem{doi:10.1126/science.1173299}
\bibinfo{author}{Barab\'{a}si, A.-L.}
\newblock \bibinfo{title}{Scale-free networks: A decade and beyond}.
\newblock \emph{\bibinfo{journal}{Science}} \textbf{\bibinfo{volume}{325}},
  \bibinfo{pages}{412--413} (\bibinfo{year}{2009}).
\newblock
  \urlprefix\url{https://www.science.org/doi/abs/10.1126/science.1173299}.

\bibitem{erdHos1960evolution}
\bibinfo{author}{Erd{\H{o}}s, P.}, \bibinfo{author}{R{\'e}nyi, A.}
  \emph{et~al.}
\newblock \bibinfo{title}{On the evolution of random graphs}.
\newblock \emph{\bibinfo{journal}{Publ. Math. Inst. Hung. Acad. Sci}}
  \textbf{\bibinfo{volume}{5}}, \bibinfo{pages}{17--60} (\bibinfo{year}{1960}).

\bibitem{Watts1998}
\bibinfo{author}{Watts, D.~J.} \& \bibinfo{author}{Strogatz, S.~H.}
\newblock \bibinfo{title}{Collective dynamics of `small-world' networks}.
\newblock \emph{\bibinfo{journal}{Nature}} \textbf{\bibinfo{volume}{393}},
  \bibinfo{pages}{440--442} (\bibinfo{year}{1998}).
\newblock \urlprefix\url{https://doi.org/10.1038/30918}.

\bibitem{Barrat2000}
\bibinfo{author}{Barrat, A.} \& \bibinfo{author}{Weigt, M.}
\newblock \bibinfo{title}{On the properties of small-world network models}.
\newblock \emph{\bibinfo{journal}{The European Physical Journal B - Condensed
  Matter and Complex Systems}} \textbf{\bibinfo{volume}{13}},
  \bibinfo{pages}{547--560} (\bibinfo{year}{2000}).
\newblock \urlprefix\url{https://doi.org/10.1007/s100510050067}.

\bibitem{Almaas2006}
\bibinfo{author}{Almaas, E.} \& \bibinfo{author}{Barab{\'a}si, A.-L.}
\newblock \emph{\bibinfo{title}{Power Laws in Biological Networks}},
  \bibinfo{pages}{1--11} (\bibinfo{publisher}{Springer US},
  \bibinfo{address}{Boston, MA}, \bibinfo{year}{2006}).

\bibitem{10.2307/2786545}
\bibinfo{author}{Travers, J.} \& \bibinfo{author}{Milgram, S.}
\newblock \bibinfo{title}{An experimental study of the small world problem}.
\newblock \emph{\bibinfo{journal}{Sociometry}} \textbf{\bibinfo{volume}{32}},
  \bibinfo{pages}{425--443} (\bibinfo{year}{1969}).
\newblock \urlprefix\url{http://www.jstor.org/stable/2786545}.

\bibitem{BLUM2005353}
\bibinfo{author}{Blum, C.}
\newblock \bibinfo{title}{Ant colony optimization: Introduction and recent
  trends}.
\newblock \emph{\bibinfo{journal}{Physics of Life Reviews}}
  \textbf{\bibinfo{volume}{2}}, \bibinfo{pages}{353--373}
  (\bibinfo{year}{2005}).
\newblock
  \urlprefix\url{https://www.sciencedirect.com/science/article/pii/S1571064505000333}.

\bibitem{Frisch+1993}
\bibinfo{author}{von Frisch, K.}
\newblock \emph{\bibinfo{title}{The Dance Language and Orientation of Bees}}
  (\bibinfo{publisher}{Harvard University Press}, \bibinfo{address}{Cambridge,
  MA and London, England}, \bibinfo{year}{1993}).
\newblock \urlprefix\url{https://doi.org/10.4159/harvard.9780674418776}.

\bibitem{mazur2018simulation}
\bibinfo{author}{Mazur, F.} \& \bibinfo{author}{Schreckenberg, M.}
\newblock \bibinfo{title}{Simulation and optimization of ground traffic on
  airports using cellular automata}.
\newblock \emph{\bibinfo{journal}{Collect. Dyn}} \textbf{\bibinfo{volume}{3}},
  \bibinfo{pages}{1--22} (\bibinfo{year}{2018}).

\bibitem{Tsuzuki_2020}
\bibinfo{author}{Tsuzuki, S.}, \bibinfo{author}{Yanagisawa, D.} \&
  \bibinfo{author}{Nishinari, K.}
\newblock \bibinfo{title}{Throughput reduction on an air-ground transport
  system by the simultaneous effect of multiple traveling routes equipped with
  parking sites}.
\newblock \emph{\bibinfo{journal}{Journal of Physics Communications}}
  \textbf{\bibinfo{volume}{4}}, \bibinfo{pages}{055009} (\bibinfo{year}{2020}).
\newblock \urlprefix\url{https://doi.org/10.1088/2399-6528/ab90c3}.

\bibitem{9878328}
\bibinfo{author}{Kawagoe, Y.}, \bibinfo{author}{Chino, R.},
  \bibinfo{author}{Tsuzuki, S.}, \bibinfo{author}{Itoh, E.} \&
  \bibinfo{author}{Okabe, T.}
\newblock \bibinfo{title}{Analyzing stochastic features in airport surface
  traffic flow using cellular automaton: Tokyo international airport}.
\newblock \emph{\bibinfo{journal}{IEEE Access}} \textbf{\bibinfo{volume}{10}},
  \bibinfo{pages}{95344--95355} (\bibinfo{year}{2022}).

\bibitem{10.5555/1102024}
\bibinfo{author}{Neumann, J.~V.} \& \bibinfo{author}{Burks, A.~W.}
\newblock \emph{\bibinfo{title}{Theory of Self-Reproducing Automata}}
  (\bibinfo{publisher}{University of Illinois Press}, \bibinfo{address}{USA},
  \bibinfo{year}{1966}).

\bibitem{RevModPhys.55.601}
\bibinfo{author}{Wolfram, S.}
\newblock \bibinfo{title}{Statistical mechanics of cellular automata}.
\newblock \emph{\bibinfo{journal}{Rev. Mod. Phys.}}
  \textbf{\bibinfo{volume}{55}}, \bibinfo{pages}{601--644}
  (\bibinfo{year}{1983}).
\newblock \urlprefix\url{https://link.aps.org/doi/10.1103/RevModPhys.55.601}.

\bibitem{9238072}
\bibinfo{author}{Wang, Z.}, \bibinfo{author}{Wang, Z.} \&
  \bibinfo{author}{Chen, C.}
\newblock \bibinfo{title}{A coordinated multiagent reinforcement learning
  method using chicken game}.
\newblock In \emph{\bibinfo{booktitle}{2020 IEEE International Conference on
  Networking, Sensing and Control (ICNSC)}}, \bibinfo{pages}{1--6}
  (\bibinfo{year}{2020}).

\bibitem{10.5555/3463952.3464013}
\bibinfo{author}{ElSayed-Aly, I.} \emph{et~al.}
\newblock \bibinfo{title}{Safe multi-agent reinforcement learning via
  shielding}.
\newblock In \emph{\bibinfo{booktitle}{Proceedings of the 20th International
  Conference on Autonomous Agents and MultiAgent Systems}}, AAMAS '21,
  \bibinfo{pages}{483^^e2^^80^^93491} (\bibinfo{publisher}{International
  Foundation for Autonomous Agents and Multiagent Systems},
  \bibinfo{address}{Richland, SC}, \bibinfo{year}{2021}).

\bibitem{10.1371/journal.pone.0013701}
\bibinfo{author}{van Wijk, B. C.~M.}, \bibinfo{author}{Stam, C.~J.} \&
  \bibinfo{author}{Daffertshofer, A.}
\newblock \bibinfo{title}{Comparing brain networks of different size and
  connectivity density using graph theory}.
\newblock \emph{\bibinfo{journal}{PLOS ONE}} \textbf{\bibinfo{volume}{5}},
  \bibinfo{pages}{1--13} (\bibinfo{year}{2010}).
\newblock \urlprefix\url{https://doi.org/10.1371/journal.pone.0013701}.

\bibitem{SciPyProceedings_11}
\bibinfo{author}{Hagberg, A.~A.}, \bibinfo{author}{Schult, D.~A.} \&
  \bibinfo{author}{Swart, P.~J.}
\newblock \bibinfo{title}{Exploring network structure, dynamics, and function
  using networkx}.
\newblock In \bibinfo{editor}{Varoquaux, G.}, \bibinfo{editor}{Vaught, T.} \&
  \bibinfo{editor}{Millman, J.} (eds.) \emph{\bibinfo{booktitle}{Proceedings of
  the 7th Python in Science Conference}}, \bibinfo{pages}{11 -- 15}
  (\bibinfo{address}{Pasadena, CA USA}, \bibinfo{year}{2008}).

\bibitem{LI2017350}
\bibinfo{author}{Li, Y.}, \bibinfo{author}{Shang, Y.} \& \bibinfo{author}{Yang,
  Y.}
\newblock \bibinfo{title}{Clustering coefficients of large networks}.
\newblock \emph{\bibinfo{journal}{Information Sciences}}
  \textbf{\bibinfo{volume}{382-383}}, \bibinfo{pages}{350--358}
  (\bibinfo{year}{2017}).
\newblock
  \urlprefix\url{https://www.sciencedirect.com/science/article/pii/S0020025516320527}.

\bibitem{WU2000597}
\bibinfo{author}{Wu, Q.}, \bibinfo{author}{Popovi^^c4^^87, D.} \&
  \bibinfo{author}{Hill, D.}
\newblock \bibinfo{title}{Avoiding sustained oscillations in power systems with
  tap changing transformers}.
\newblock \emph{\bibinfo{journal}{International Journal of Electrical Power \&
  Energy Systems}} \textbf{\bibinfo{volume}{22}}, \bibinfo{pages}{597--605}
  (\bibinfo{year}{2000}).
\newblock
  \urlprefix\url{https://www.sciencedirect.com/science/article/pii/S0142061500000259}.

\bibitem{Balaban2018}
\bibinfo{author}{Balaban, V.}, \bibinfo{author}{Lim, S.},
  \bibinfo{author}{Gupta, G.}, \bibinfo{author}{Boedicker, J.} \&
  \bibinfo{author}{Bogdan, P.}
\newblock \bibinfo{title}{Quantifying emergence and self-organisation of
  enterobacter cloacae microbial communities}.
\newblock \emph{\bibinfo{journal}{Scientific Reports}}
  \textbf{\bibinfo{volume}{8}}, \bibinfo{pages}{12416} (\bibinfo{year}{2018}).
\newblock \urlprefix\url{https://doi.org/10.1038/s41598-018-30654-9}.

\bibitem{Yang2021}
\bibinfo{author}{Yang, R.}, \bibinfo{author}{Sala, F.} \&
  \bibinfo{author}{Bogdan, P.}
\newblock \bibinfo{title}{Hidden network generating rules from partially
  observed complex networks}.
\newblock \emph{\bibinfo{journal}{Communications Physics}}
  \textbf{\bibinfo{volume}{4}}, \bibinfo{pages}{199} (\bibinfo{year}{2021}).
\newblock \urlprefix\url{https://doi.org/10.1038/s42005-021-00701-5}.

\bibitem{Xiao2021}
\bibinfo{author}{Xiao, X.}, \bibinfo{author}{Chen, H.} \&
  \bibinfo{author}{Bogdan, P.}
\newblock \bibinfo{title}{Deciphering the generating rules and functionalities
  of complex networks}.
\newblock \emph{\bibinfo{journal}{Scientific Reports}}
  \textbf{\bibinfo{volume}{11}}, \bibinfo{pages}{22964} (\bibinfo{year}{2021}).
\newblock \urlprefix\url{https://doi.org/10.1038/s41598-021-02203-4}.

\bibitem{Koorehdavoudi2016}
\bibinfo{author}{Koorehdavoudi, H.} \& \bibinfo{author}{Bogdan, P.}
\newblock \bibinfo{title}{A statistical physics characterization of the complex
  systems dynamics: Quantifying complexity from spatio-temporal interactions}.
\newblock \emph{\bibinfo{journal}{Scientific Reports}}
  \textbf{\bibinfo{volume}{6}}, \bibinfo{pages}{27602} (\bibinfo{year}{2016}).
\newblock \urlprefix\url{https://doi.org/10.1038/srep27602}.

\bibitem{Callen:450289}
\bibinfo{author}{Callen, H.~B.}
\newblock \emph{\bibinfo{title}{{Thermodynamics and an introduction to
  thermostatistics; 2nd ed.}}} (\bibinfo{publisher}{Wiley},
  \bibinfo{address}{New York, NY}, \bibinfo{year}{1985}).

\bibitem{doi:10.1080/10610270108027472}
\bibinfo{author}{Messina, M.~T.} \emph{et~al.}
\newblock \bibinfo{title}{Herringbone infinite networks formed by terpyridine
  and haloperfluoroarene modules}.
\newblock \emph{\bibinfo{journal}{Supramolecular Chemistry}}
  \textbf{\bibinfo{volume}{12}}, \bibinfo{pages}{405--410}
  (\bibinfo{year}{2001}).
\newblock \urlprefix\url{https://doi.org/10.1080/10610270108027472}.

\bibitem{doi:10.1021/cg034199d}
\bibinfo{author}{McManus, G.~J.}, \bibinfo{author}{Wang, Z.} \&
  \bibinfo{author}{Zaworotko, M.~J.}
\newblock \bibinfo{title}{Suprasupermolecular chemistry: Infinite networks from
  nanoscale metal-organic building blocks}.
\newblock \emph{\bibinfo{journal}{Crystal Growth {\&} Design}}
  \textbf{\bibinfo{volume}{4}}, \bibinfo{pages}{11--13} (\bibinfo{year}{2004}).
\newblock \urlprefix\url{https://doi.org/10.1021/cg034199d}.

\bibitem{doi:10.1021/acs.inorgchem.6b01199}
\bibinfo{author}{Fern^^c3^^a1ndez~de Luis, R.}, \bibinfo{author}{Larrea,
  E.~S.}, \bibinfo{author}{Orive, J.}, \bibinfo{author}{Lezama, L.} \&
  \bibinfo{author}{Arriortua, M.~I.}
\newblock \bibinfo{title}{Commensurate superstructure of the
  {Cu(NO3)(H2O)}(htae)(bpy) coordination polymer: An example of 2d
  hydrogen-bonding networks as magnetic exchange pathway}.
\newblock \emph{\bibinfo{journal}{Inorganic Chemistry}}
  \textbf{\bibinfo{volume}{55}}, \bibinfo{pages}{11662--11675}
  (\bibinfo{year}{2016}).
\newblock \urlprefix\url{https://doi.org/10.1021/acs.inorgchem.6b01199}.
\newblock \bibinfo{note}{PMID: 27805389}.

\bibitem{doi:10.1021/nn502149u}
\bibinfo{author}{Guerette, P.~A.} \emph{et~al.}
\newblock \bibinfo{title}{Nanoconfined $\beta$-sheets mechanically reinforce
  the supra-biomolecular network of robust squid sucker ring teeth}.
\newblock \emph{\bibinfo{journal}{ACS Nano}} \textbf{\bibinfo{volume}{8}},
  \bibinfo{pages}{7170--7179} (\bibinfo{year}{2014}).
\newblock \urlprefix\url{https://doi.org/10.1021/nn502149u}.
\newblock \bibinfo{note}{PMID: 24911543}.

\bibitem{Chung2010}
\bibinfo{author}{Chung, F.} \& \bibinfo{author}{Zhao, W.}
\newblock \emph{\bibinfo{title}{PageRank and Random Walks on Graphs}},
  \bibinfo{pages}{43--62} (\bibinfo{publisher}{Springer Berlin Heidelberg},
  \bibinfo{address}{Berlin, Heidelberg}, \bibinfo{year}{2010}).
\newblock \urlprefix\url{https://doi.org/10.1007/978-3-642-13580-4_3}.

\bibitem{LAI20102443}
\bibinfo{author}{Lai, D.}, \bibinfo{author}{Lu, H.} \&
  \bibinfo{author}{Nardini, C.}
\newblock \bibinfo{title}{Finding communities in directed networks by pagerank
  random walk induced network embedding}.
\newblock \emph{\bibinfo{journal}{Physica A: Statistical Mechanics and its
  Applications}} \textbf{\bibinfo{volume}{389}}, \bibinfo{pages}{2443--2454}
  (\bibinfo{year}{2010}).
\newblock
  \urlprefix\url{https://www.sciencedirect.com/science/article/pii/S0378437110001391}.

\bibitem{NAGATANI201866}
\bibinfo{author}{Nagatani, T.}, \bibinfo{author}{Ichinose, G.} \&
  \bibinfo{author}{ichi Tainaka, K.}
\newblock \bibinfo{title}{Epidemics of random walkers in metapopulation model
  for complete, cycle, and star graphs}.
\newblock \emph{\bibinfo{journal}{Journal of Theoretical Biology}}
  \textbf{\bibinfo{volume}{450}}, \bibinfo{pages}{66--75}
  (\bibinfo{year}{2018}).
\newblock
  \urlprefix\url{https://www.sciencedirect.com/science/article/pii/S0022519318302078}.

\bibitem{ABBAS2019193}
\bibinfo{author}{Abbas, A.} \emph{et~al.}
\newblock \bibinfo{title}{Quasi-periodic patterns contribute to functional
  connectivity in the brain}.
\newblock \emph{\bibinfo{journal}{NeuroImage}} \textbf{\bibinfo{volume}{191}},
  \bibinfo{pages}{193--204} (\bibinfo{year}{2019}).
\newblock
  \urlprefix\url{https://www.sciencedirect.com/science/article/pii/S1053811919300825}.

\bibitem{Belloy2018}
\bibinfo{author}{Belloy, M.~E.} \emph{et~al.}
\newblock \bibinfo{title}{Quasi-periodic patterns of neural activity improve
  classification of alzheimer-disease in mice}.
\newblock \emph{\bibinfo{journal}{Scientific Reports}}
  \textbf{\bibinfo{volume}{8}}, \bibinfo{pages}{10024} (\bibinfo{year}{2018}).
\newblock \urlprefix\url{https://doi.org/10.1038/s41598-018-28237-9}.

\bibitem{ABBAS2019101653}
\bibinfo{author}{Abbas, A.}, \bibinfo{author}{Bassil, Y.} \&
  \bibinfo{author}{Keilholz, S.}
\newblock \bibinfo{title}{Quasi-periodic patterns of brain activity in
  individuals with attention-deficit/hyperactivity disorder}.
\newblock \emph{\bibinfo{journal}{NeuroImage: Clinical}}
  \textbf{\bibinfo{volume}{21}}, \bibinfo{pages}{101653}
  (\bibinfo{year}{2019}).
\newblock
  \urlprefix\url{https://www.sciencedirect.com/science/article/pii/S2213158219300038}.

\bibitem{Kyriakis2020}
\bibinfo{author}{Kyriakis, P.}, \bibinfo{author}{Pequito, S.} \&
  \bibinfo{author}{Bogdan, P.}
\newblock \bibinfo{title}{On the effects of memory and topology on the
  controllability of complex dynamical networks}.
\newblock \emph{\bibinfo{journal}{Scientific Reports}}
  \textbf{\bibinfo{volume}{10}}, \bibinfo{pages}{17346} (\bibinfo{year}{2020}).
\newblock \urlprefix\url{https://doi.org/10.1038/s41598-020-74269-5}.

\bibitem{8815074}
\bibinfo{author}{Gupta, G.}, \bibinfo{author}{Pequito, S.} \&
  \bibinfo{author}{Bogdan, P.}
\newblock \bibinfo{title}{Learning latent fractional dynamics with unknown
  unknowns}.
\newblock In \emph{\bibinfo{booktitle}{2019 American Control Conference
  (ACC)}}, \bibinfo{pages}{217--222} (\bibinfo{year}{2019}).

\bibitem{Bu2023}
\bibinfo{author}{Bu, F.}, \bibinfo{author}{Kang, S.} \& \bibinfo{author}{Shin,
  K.}
\newblock \bibinfo{title}{Interplay between topology and edge weights in
  real-world graphs: concepts, patterns, and an algorithm}.
\newblock \emph{\bibinfo{journal}{Data Mining and Knowledge Discovery}}
  \textbf{\bibinfo{volume}{37}}, \bibinfo{pages}{2139--2191}
  (\bibinfo{year}{2023}).
\newblock \urlprefix\url{https://doi.org/10.1007/s10618-023-00940-w}.

\bibitem{Xue2019}
\bibinfo{author}{Xue, Y.} \& \bibinfo{author}{Bogdan, P.}
\newblock \bibinfo{title}{Reconstructing missing complex networks against
  adversarial interventions}.
\newblock \emph{\bibinfo{journal}{Nature Communications}}
  \textbf{\bibinfo{volume}{10}}, \bibinfo{pages}{1738} (\bibinfo{year}{2019}).
\newblock \urlprefix\url{https://doi.org/10.1038/s41467-019-09774-x}.

\end{thebibliography}

\clearpage

\begin{figure*}[t]
\vspace{-87.8cm}
\includegraphics[width=4.35\textwidth, clip, bb= 0 0 2478 3586]{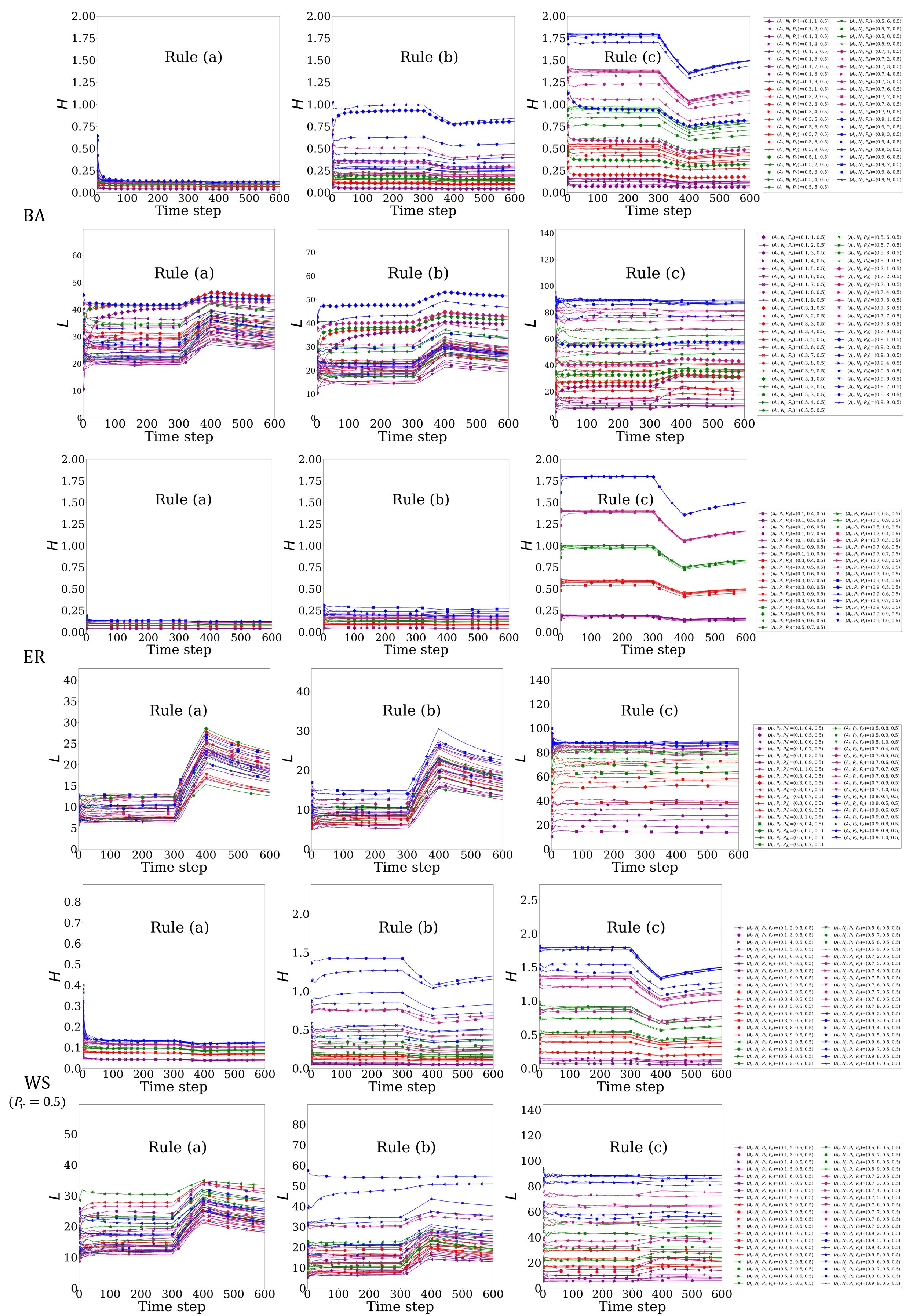}
\caption{Time-dependence of the hunting rate $H$ and the imbalance rate $L$ during the simulation where agents follow Rules (a)--(c) on BA, ER, and WS network models with a functionality of disconnecting nodes in a certain time period of 300 to 400 time steps out of 600 time steps.}
\label{fig:Figure10}
\end{figure*}

\end{document}